\documentclass[aps,pre,showpacs,showkeys,amssymb,twocolumn]{revtex4}

\usepackage{bm}
\usepackage{graphicx}
\usepackage{amsfonts}
\usepackage{amsmath}

\begin{document}

\title{Models with symmetry breaking phase transitions triggered by dumbbell-shaped-equipotential surfaces}

\author{Fabrizio Baroni}

\email{f.baroni@ifac.cnr.it, baronifab@libero.it}
\affiliation{IFAC-CNR Institute of applied physics "Nello Carrara", Via Madonna del Piano 10, I-50019 Sesto Fiorentino (FI), Italy}

\date{\today}

\begin{abstract}
In some recent papers some sufficiency conditions for the occurrence of a $\mathbb{Z}_2$-symmetry breaking phase transition ($\mathbb{Z}_2$-SBPT) have been showed starting from geometric-topological concepts of potential energy landscapes. In particular, a $\mathbb{Z}_2$-SBPT can be triggered by double-well potentials, or in an equivalent way, by dumbbell-shaped equipotential surfaces. In this paper we introduce two models with a $\mathbb{Z}_2$-SBPT which, due to their essential feature, show in the clearest way the generating-mechanism of a $\mathbb{Z}_2$-SBPT above mentioned. These models, despite they cannot be considered physical models, have all the features of such models with the same kind of SBPT. At the end of the paper, the $\phi^4$ model is revisited in the light of this approach. In particular, the landscape of one of the model introduced here is turned out to be equivalent to that of the mean-field $\phi^4$ model in a simplified version.  
\end{abstract}



\maketitle


\section{Introduction}

Phase transition (PT) are very common in nature. They are sudden changes of the macroscopic behavior of a natural system composed by many interacting parts occurring while an external parameter is smoothly varied. PTs are an example of emergent behavior, i.e., of a collective properties having no direct counterpart in the dynamics or structure of individual atoms \cite{lebowitz}. The successful description of PTs starting from the properties of the interactions between the components of the system is one of the major achievements of equilibrium statistical mechanics.

From a statistical-mechanical viewpoint, in the canonical ensemble, describing a system at constant temperature $T$, a PT occurs at special $T$-values called transition points, where thermodynamic quantities such as pressure, magnetization, or heat capacity, are non-analytic-$T$ functions; these points are the boundaries between different phases of the system. PTs are strictly related to the phenomenon of spontaneous symmetry breaking (SB). For example, in a natural magnet below the Curie temperature the $0(3)$ symmetry is spontaneous broken. This is witnessed by the occurrence of a non-vanishing spontaneous magnetization. In this paper we mostly consider the origin of spontaneous symmetry breaking, and secondarily the origin of non-analytic points in the thermodynamic functions.

Despite great achievements in our understanding of PTs, yet, the situation is not completely satisfactory. For example, while necessary conditions for the presence of a PT can be found, nothing general is known about sufficient conditions, apart some particular cases \cite{bc,b_2}: no general procedure is at hand to tell if a system where a PT is not ruled out from the beginning does have or not such a transition without computing the partition function $Z$. This might indicate that our deep understanding of this phenomenon is still incomplete. 

These considerations motivate a study of PTs based on alternative approaches. One of them is the geometric-topological approach based on the study of energy potential landscapes. In particular, equipotential surfaces, i.e., potential level sets ($v$-level sets), have gained a great importance inside this approach. In addition to the study of the $v$-level sets, the study of the critical points has also assumed considerable importance. These ideas have been developed and discussed in many recent papers \cite{acprz,b_0,ccp2,ckn,gss,gm,k,rs,nc,cnn,km,mhk,nkmc,dgppnfv,gfp,gfp1,gfp2,pgfcp,cccp,ccp,ccp1,ck,ck1,cfsp,fp,fp1,hk1,ks,kss1,kss,mk}.

In particular, in Ref. \cite{b_3} it has been stated a links between the occurrence of a $\mathbb{Z}_2$-SBPT and \emph{dumbbell-shaped} $v$-level sets. Intuitively, a $v$-level set is said to be dumbbell-shaped when it is made up of two major components connected by a shrink neck. Something like this SBPT generating-mechanism has been put forward also in Refs. \cite{gfp,gfp1}. According to this framework, a spontaneous $\mathbb{Z}_2$-SB is entailed by dumbbell-shaped $v$-level sets. The thermodynamic critical potential $\langle v\rangle_c$ is in correspondence of a particular $v_c$-level set which can be said critical in the sense that it is the boundary between the dumbbell $v$-level sets for $v<v_c$ and the non-dumbbell ones for $v>v_c$. An advantage with respect to the traditional definition of SBPTs is that this definition holds for finite $N$, so that it is not necessary resorting to the thermodynamic limit in order to define a SBPT. Since in the last decades many examples of transitional phenomena in systems far form the thermodynamic limit have been found (e.g., in nuclei, atomic clusters, biopolymers, superconductivity, superfluidity), a description of SBPTs valid also for finite systems would be desirable. 

In this paper we will introduce two models showing a $\mathbb{Z}_2$-SBPT which illustrate in the clearest way the generating-mechanism based on the concept of dumbbell-shaped $v$-level sets. Such sets are generated in turn by double-well potentials. These models do not describe any physical system, so that their usefulness is for giving hints about physical models, and for didactic purposes. Despite this, one of the two models is very close to a physical model, i.e., the well known mean-field $\phi^4$ model with a suitable simplification. 

The structure of the paper is as follows. In Sec. \ref{framework} we will introduce the framework of the geometric-topological approach to SBPTs in the canonical treatment. In Sec. \ref{rev} we will build a model with non-smooth potential. In Sec. \ref{revs} we will derive from that another model with smooth potential. The potential landscape of this model is characterized by the presence of three stationary points only. Finally, in Sec. \ref{phi4} we will revisit the mean-field $\phi^4$ model in the light of the scenario of dumbbell-shaped $v$-level sets and make a comparison with the model with smooth potential introduced here.

\section{Framework of the geometric-topological approach to SBPTs}
\label{framework}

Hereafter, we will refer to the canonical treatment, although the dumbbell-shaped $v$-level set approach can be extended to the microcanonical one.

Consider an $N$ degrees of freedom system with Hamiltonian given by
\begin{equation}
H(\textbf{p},\textbf{q})=T+V=\sum_{i=1}^N \frac{p_i^2}{2}+V(\textbf{q}).
\end{equation}
Let $M\subseteq\mathbb{R}^N$ be the configuration space. The partition function is by definition
\begin{eqnarray}
Z(\beta,N)=\int_{\mathbb{R}^N\times M} \rm d\mathbf{p}\,\rm d\mathbf{q}\,e^{-\beta H(\mathbf{p},\mathbf{q})}=\nonumber
\\
=\int_{\mathbb{R}^N} \rm d\mathbf{p}\,e^{-\beta\sum_{i=1}^N \frac{p_i^2}{2}}\int_M \rm d\mathbf{q}\,e^{-\beta V(\textbf{q})}=Z_{kin}Z_c
\end{eqnarray}
where $\beta=1/T$ (in unit $k_B=1$), $Z_{kin}$ is the kinetic part of $Z$, and $Z_c$ is the configurational part. In order to develop what follows we assume the potential to be lower bounded, thus 
$Z_c$ can be written according to the coarea formula \cite{fr} as follows 
\begin{eqnarray}
Z_c=N\int_{v_{min}}^{+\infty}\rm dv\,e^{-\beta Nv}\int_{\Sigma_{v,N}}\frac{\rm d\Sigma}{\left\|\nabla V\right\|},
\label{zc}
\end{eqnarray}
where $v=V/N$ is the potential density, and the $\Sigma_{v,N}$'s are the $v$-level sets defined as
\begin{equation}
\Sigma_{v,N}=\{\textbf{q}\in M: v(\textbf{q})=v\}.
\label{sigmav}
\end{equation}
The set of the $\Sigma_{v,N}$'s is a foliation of configuration space $M$ while varying $v$ between $v_{min}$ and $+\infty$. The $\Sigma_{v,N}$'s are very important submanifolds of $M$ because as $N\rightarrow\infty$ the canonical statistic measure shrinks around $\Sigma_{\langle v\rangle(T),N}$, where $\langle v\rangle(T)$ is the average potential density. Thus, $\Sigma_{\langle v\rangle(T),N}$ becomes the most probably accessible $v$-level set by the representative point of the system. This fact may have significant consequences on the symmetries of the system because the ergodicity may be broken by the mechanisms pointed out in Refs. \cite{bc,b_3}.

We can make the same considerations about $Z_{kin}$, but the related submanifolds $\Sigma_{e,N}$, where $e=E/N$ is the kinetic energy density, are all trivially $N$-spheres, thus they cannot affect the symmetry properties of the system. Furthermore, $Z_{kin}$ is analytic at any $T$ in the thermodynamic limit, so that it cannot entail any loss of analyticity in $Z$. For the above considerations, hereafter we will consider $Z_c$ alone, so as for the thermodynamic functions.

\section{The revolution model by the 'double-well method'}
\label{rev}

In Ref. \cite{bc,b_2} a sufficiency condition for the occurrence of a $\mathbb{Z}_2$-SBPT has been proven. This condition is a double-well potential with two global minima separated by a gap proportional to $N$. Here, we will apply this result ('double-well method') to build a toy model with a $\mathbb{Z}_2$-SBPT that we will call \emph{revolution model}. 

Let $q_1,\cdots,q_{N}$ be the standard coordinate system of $\mathbb{R}^N$. The starting point is setting a new coordinate system
\begin{equation}
\left(m,\widetilde{q}_1,\cdots,\widetilde{q}_{N-1}\right),
\label{cs}
\end{equation}
where $m=1/N\sum^{N}_{i=1}q_i$. $m$ labels the points on the line orthogonal to the hyperplane at constant $m$ passing through the origin and being distant $\sqrt{N}m$ from it. $(\widetilde{q}_1,\cdots,\widetilde{q}_{N-1})$ are an orthonormal coordinate system contained in the hyperplane at constant $m$ for $m=0$ with the origin coinciding with that of the standard coordinate system.

At first, we define the potential as an $m$-function as follows
\begin{equation}
V=N(m^4-Jm^2),
\end{equation}
where $J>0$ plays the role of a coupling constant. $V$ is flat along the hyperplanes at constant $m$. $V$ has two degenerate global minima at $m=\pm \sqrt{J/2}$ of value $-NJ^2/4$, whose coordinates in the system (\ref{cs}) are $\left(\pm \sqrt{J/2},\widetilde{q}_1,\cdots,\widetilde{q}_{N-1}\right)$ for every $\widetilde{q}_i$ with $i=1,\cdots,N-1$. Furthermore, $V$ has a degenerate local maximum at $m=0$ of coordinates $\left(0,\widetilde{q}_1,\cdots,\widetilde{q}_{N-1}\right)$ for every $\widetilde{q}_i$.

\begin{figure}
	\begin{center}
		\includegraphics[width=0.5\textwidth]{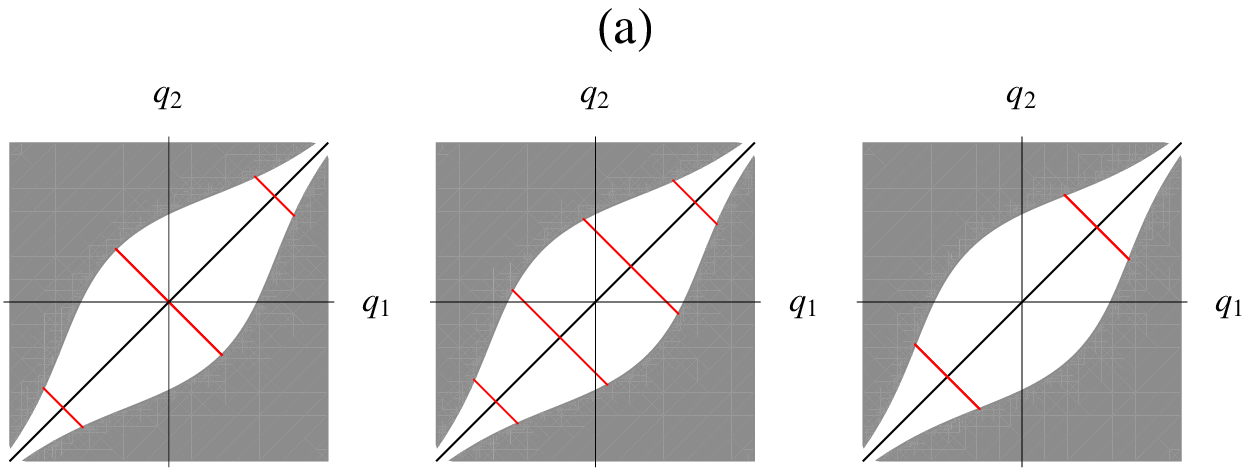}
		\includegraphics[width=0.5\textwidth]{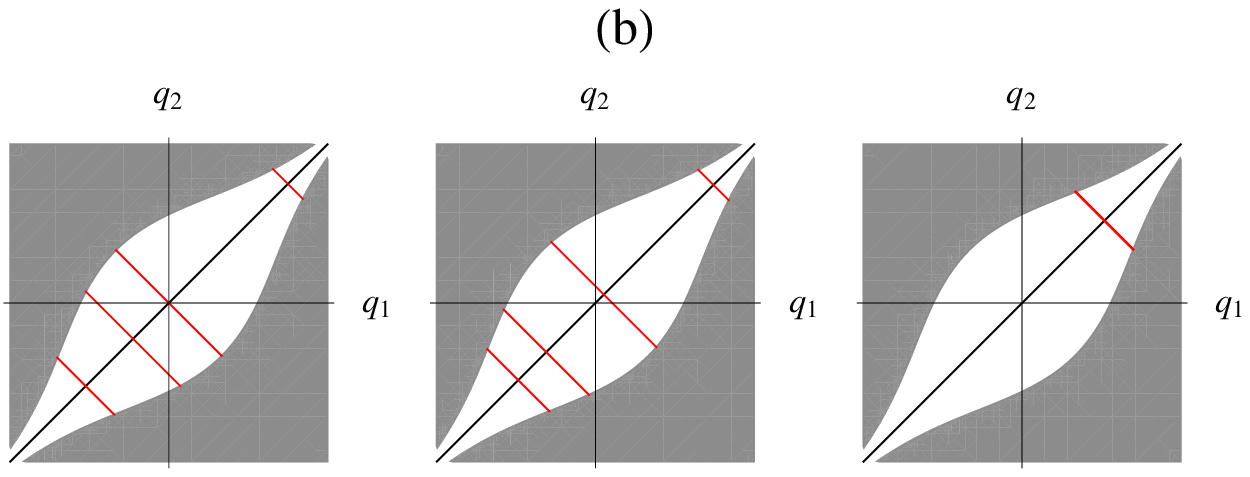}
		\caption{(a) Some $\Sigma_{v,N}$'s (red segments) for $N=2$ of the model (\ref{Vrev}) for $J=1$, external magnetic field $H=0$, and potential values $v=0, -0.1, -0.25$ from left to right, respectively. The configuration space $M$ is the white region. (b) The same of the panel (a) for $H=0.3$ and $v=0, -0.03, -0.47$, respectively.}
		\label{fig_rev_sigmav}
	\end{center}
\end{figure}

The $\Sigma_{V,N}$'s describe hyperplanes of $\mathbb{R}^N$ at constant $m$, that for convenience we define in the standard coordinate system as follows
\begin{equation}
\Sigma_{m,N}=\{\mathbf{q}\in\mathbb{R}^N: \frac{1}{N}\sum^{N}_{i=1}q_i=m\},
\label{pi}
\end{equation}
or in the coordinate system (\ref{cs})
\begin{equation}
\Sigma_{m,N}=\{(m',\widetilde{q}_1,\cdots,\widetilde{q}_{N-1})\in\mathbb{R}^N: m'=m\}.
\label{pimcs}
\end{equation}
Now, to define the double well we introduce a suitable constraint $M\subseteq\mathbb{R}^N$ which will be the configuration space. We will define it in such a way that the Taylor expansion of the entropy at fixed $m$
\begin{eqnarray}
s_N(m)&=&\frac{1}{N}\ln \mu(M\cap\Sigma_{m,N})=\nonumber
\\
&=&\frac{1}{N}\ln\left(\int_{M\cap\Sigma_{m,N}}\frac{d\Sigma}{\Vert\nabla M\Vert}\right)=\nonumber
\\
&=&\frac{1}{N}\ln \left(\sqrt{N} vol(M\cap\Sigma_{m,N})\right)
\end{eqnarray}
has a non-vanishing-second-order term. '$vol(\cdot)$' stands for the volume calculated by the standard measure on $M\cap\Sigma_{m,N}$ entailed by the embedding in $\mathbb{R}^N$. This term will compete with the second-order term of the Taylor expansion of $V$ giving rise to the critical temperature $T_c$. To attain this effect, the simplest choice is
\begin{equation}
vol(M\cap\Sigma_{m,N})=e^{-(N-1)m^2},
\label{volrev}
\end{equation}
which yields in the thermodynamic limit
\begin{equation}
s(m)=\lim_{N\to\infty}s_N(m)=-m^2.
\end{equation}
Now, the problem is to fit up these volumes to complete the definition of $M$. We can give them the shape of an $(N-1)$-ball contained in the hyperplane $\Sigma_{m,N}$ with the center belonging to the line orthogonal to $\Sigma_{m,N}$ and passing through the origin. The radius of the $(N-1)$-ball is chosen in such a way to yield the volume (\ref{volrev}). 
Summarizing, the complete definition of the potential is the following
\begin{equation}
V=\left\{\begin{array}{ll}
N(m^4-Jm^2)  \quad & \hbox{if} \quad \mathbf{q}\in M
\\
+\infty\quad  & \hbox{if}\quad  \mathbf{q}\notin M
\end{array}\right.,
\label{Vrev}
\end{equation}
where
\begin{equation}
M=\cup_{m\in\mathbb{R}}\overline{B}_R^N(m,0,\cdots,0),
\end{equation}
where $\overline{B}_R^N(m,0,\cdots,0)$ is the closed $N$-ball of radius $R$, of volume (\ref{volrev}), and with the center coordinates $(m,0,\cdots,0)$ expressed in the system (\ref{cs}). So built, $M$ has the shape of an $N$-dimensional 'spindle' of infinite length (see Fig. \ref{fig_rev_sigmav}). The potential $V$, besides the $\mathbb{Z}_2$ symmetry, has also an $O(N-1)$ symmetry (from which the name \emph{revolution model}).

\subsection{Canonical thermodynamic}
\label{rev_ct}

\begin{figure}
	\begin{center}
		\includegraphics[width=0.235\textwidth]{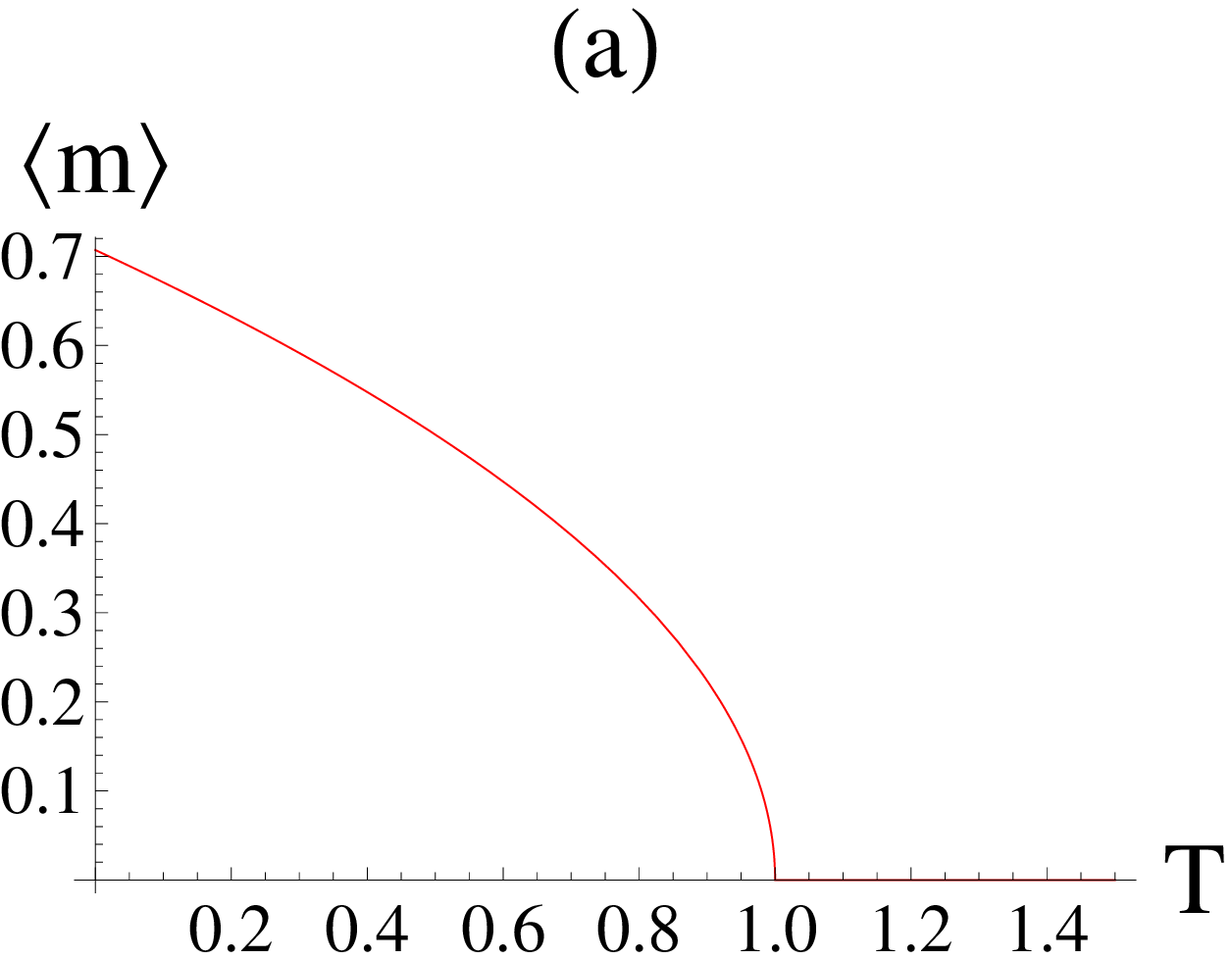}
		\includegraphics[width=0.235\textwidth]{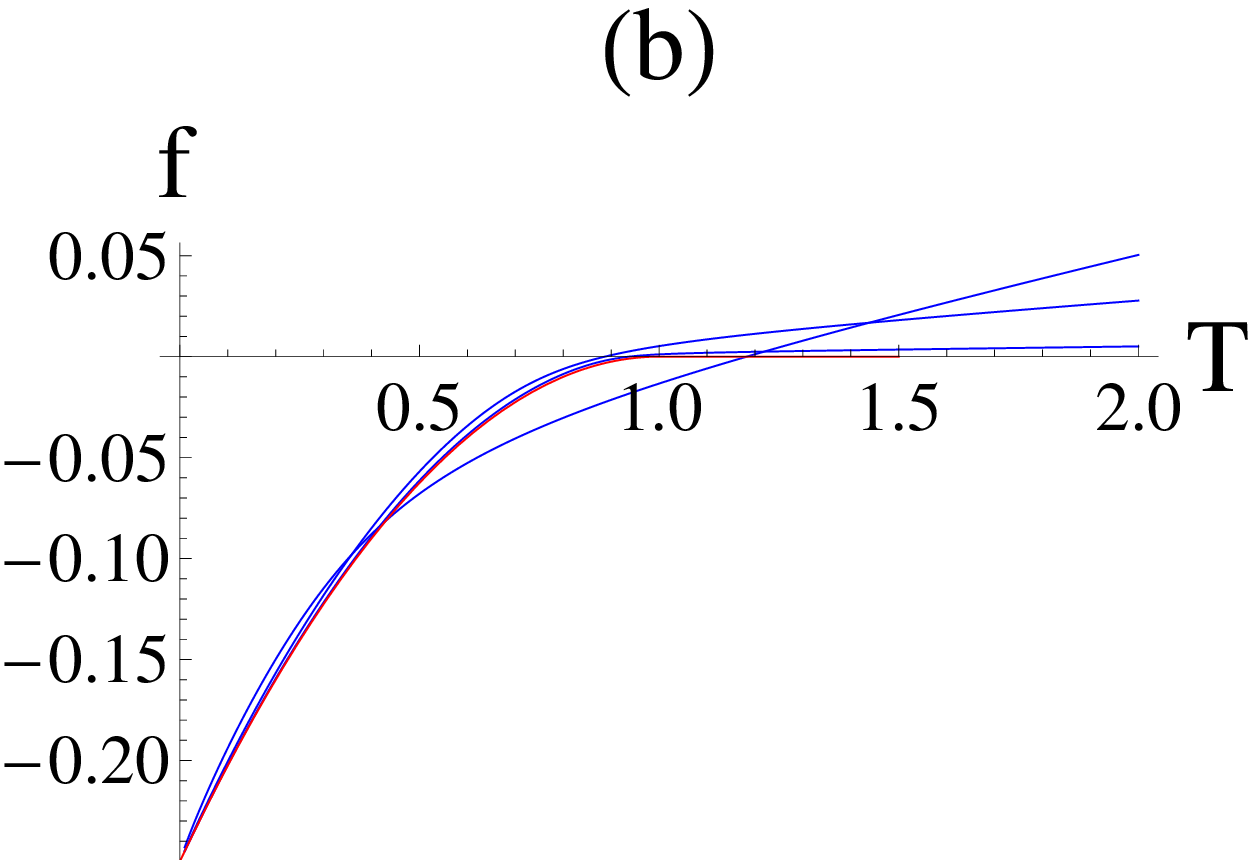}
		\includegraphics[width=0.235\textwidth]{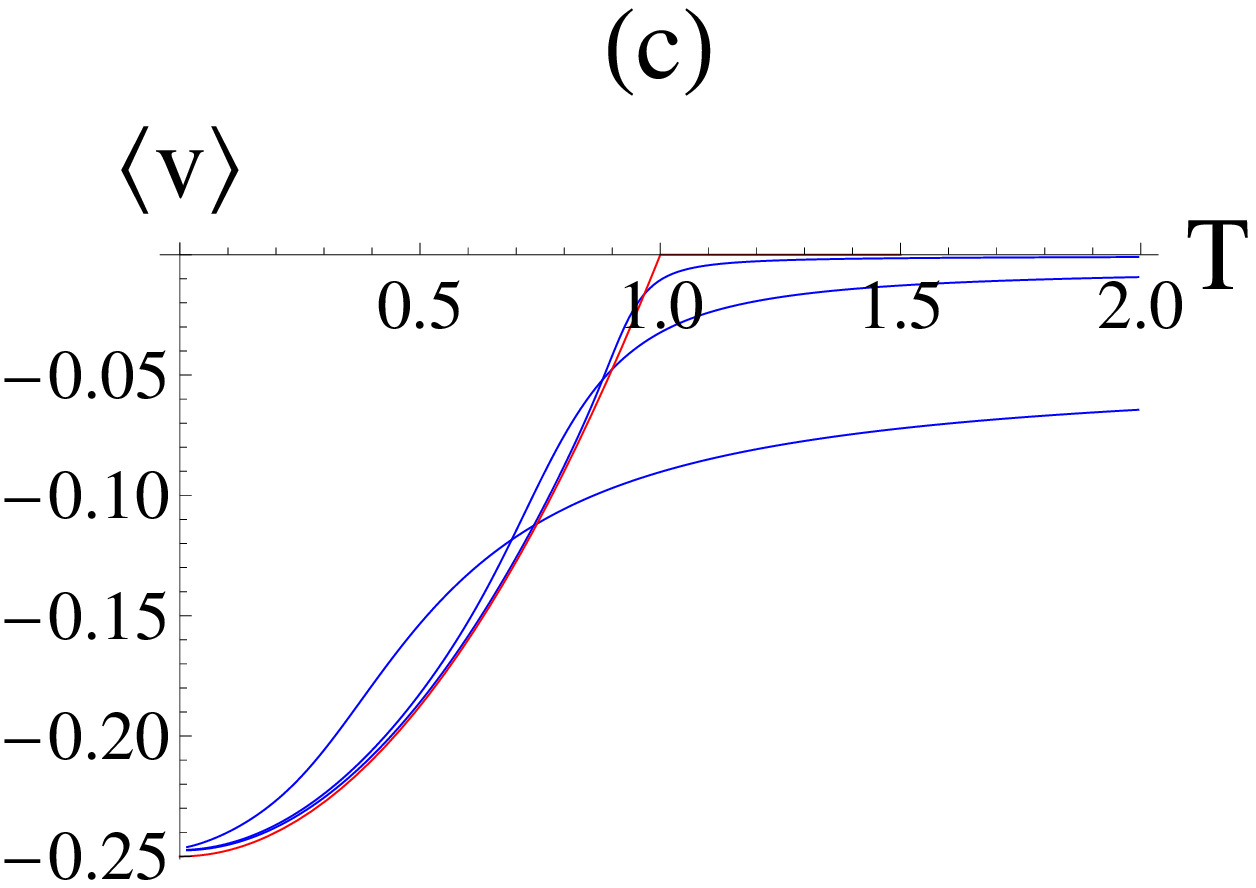}
		\includegraphics[width=0.235\textwidth]{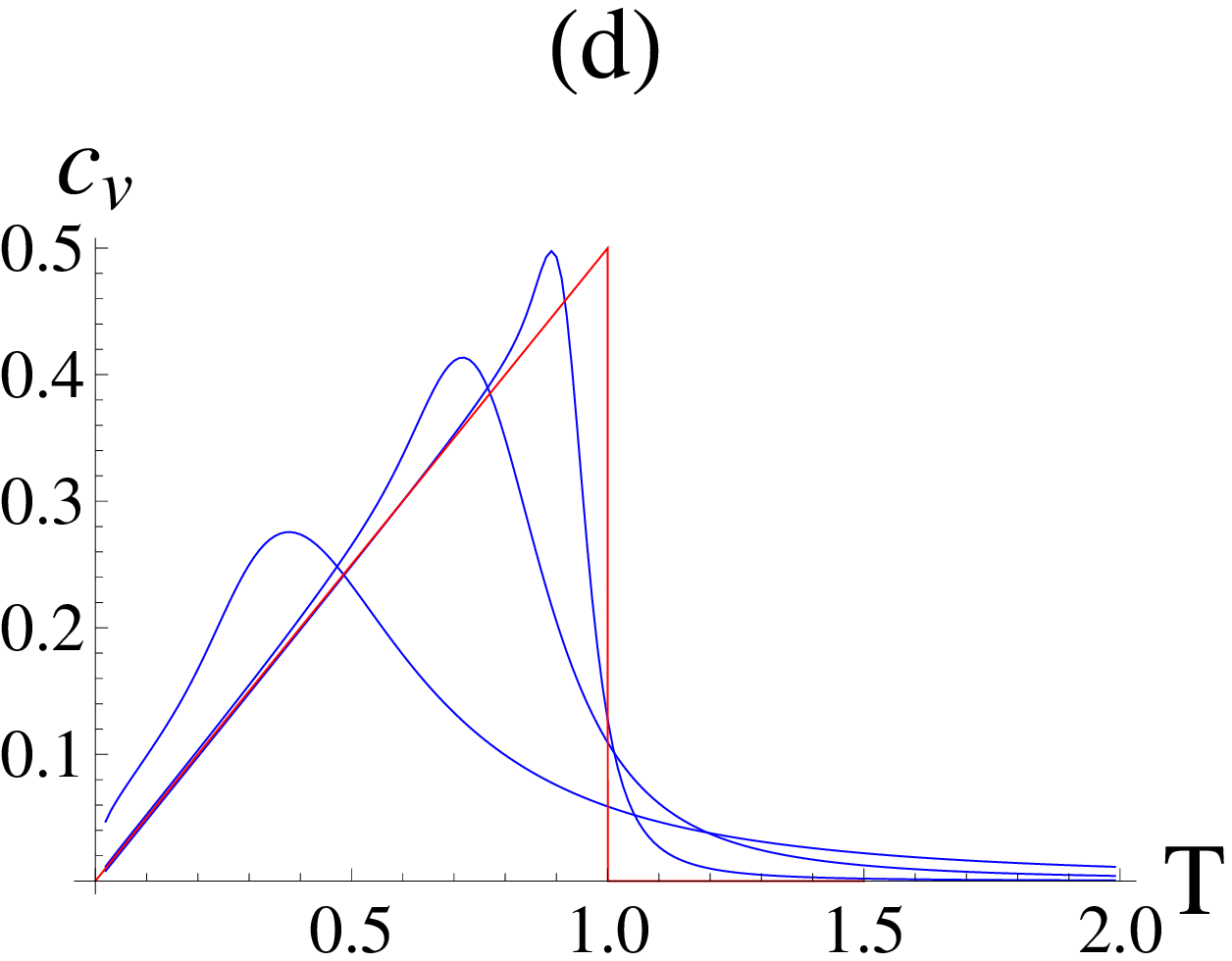}
		\caption{Model (\ref{Vrev}) for $J=1$. (a) Spontaneous magnetization, (b) free energy, (c) average potential, and (d) specific heat as functions of the temperature. The blue graphs are for $N=10, ~100,~ 1000$, while the red ones are for $N=\infty$.}
		\label{fig_rev_terfun}
	\end{center}
\end{figure}

The free energy results to be 
\begin{equation}
f=v(m)-Ts(m)=m^4+(T-J)m^2,
\label{fmT}
\end{equation}
where $v=V/N$. $T_c=J$ is the critical temperature of the system. The spontaneous magnetization is given by a minimization process of $f$ with respect to $m$, and results as follows
\begin{equation}
\left\langle m\right\rangle=\left\{\begin{array}{ll}
\pm \left(\frac{1}{2}(J-T)\right)^{\frac{1}{2}}  & \hbox{if} \quad T\le T_c
\\
0   & \hbox{if} \quad T\ge T_c
\end{array}\right..
\label{mT}
\end{equation}
The free energy, the average potential, and the specific heat  as functions of $T$ are, respectively,
\begin{equation}
f(m(T),T)=\left\{\begin{array}{ll}
-\frac{1}{4}(J-T)^2  & \hbox{if} \quad T\le T_c
\\
0   & \hbox{if} \quad T\ge T_c
\end{array}\right.,
\end{equation}
\begin{equation}
\left\langle v\right\rangle=-T^2\frac{\partial}{\partial T}\left(\frac{f}{T}\right)=\left\{\begin{array}{ll}
-\frac{1}{4}JT^2  & \hbox{if} \quad T\le T_c
\\
0   & \hbox{if} \quad T\ge T_c
\end{array}\right.,
\end{equation}
\begin{equation}
c_v=\frac{\partial\left\langle v\right\rangle}{\partial T}=\left\{\begin{array}{ll}
\frac{1}{2}T  & \hbox{if} \quad T\le T_c
\\
0   & \hbox{if} \quad T\ge T_c.
\end{array}\right..
\end{equation}
The partition function is
\begin{equation}
Z_N=\sqrt{N}\int dm\,e^{-\frac{N}{T}(m^4-m^2)-(N-1)m^2},
\end{equation}
which allows the calculation at finite $N$ of the thermodynamic functions (see Fig. \ref{fig_rev_terfun}). We avoid to report the calculations here because they are trivial. The uniform convergence toward the limit $N\to\infty$ is broken in correspondence of the critical temperature $T_c$.

\subsection{External magnetic field and critical exponents}
\label{rev_emf}

\begin{figure}
	\begin{center}
		\includegraphics[width=0.235\textwidth]{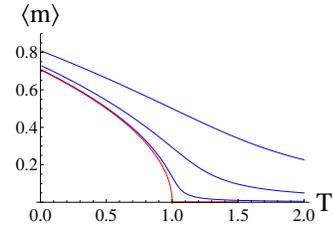}
		\caption{Model (\ref{Vrev}). Effect of an external magnetic field $H$ on the spontaneous magnetization as a function of the temperature. The blue graphs are for $H=0.01, ~0.1, ~0.5$ from the lowest to the highest, respectively. The red graph is for $H=0$.}
		\label{TmH}
	\end{center}
\end{figure}

Our purpose is to find out the critical exponent $\alpha, \beta, \gamma, \delta$ of the $\mathbb{Z}_2$-SBPT. $\alpha=0$ because the specific heat $c_v(T)$ has a finite jump at $T_c$. $\beta=1/2$ has been already found out in Sec. \ref{rev_ct} because $\left\langle m\right\rangle\propto \sqrt{T-T_c}$. 
The effect of an external magnetic field $H$ can be taken into account by the Hamiltonian interacting term
\begin{equation}
V_H=-H\sum_{i=1}^N q_i=-NHm.
\end{equation}
The free energy (\ref{fmT}) becomes
\begin{equation}
f(m,T)=m^4+(T-J)m^2-mH.
\end{equation}
By solving the third-order equation in $m$ $\partial f/\partial m=0$ we obtain the spontaneous magnetization $\left\langle m\right\rangle(H,J,T)$. The solution is trivial but quite complicated and we prefer not to report it here (see Fig. \ref{TmH} for a plot).
By inserting $T=T_c=J$ in $\left\langle m\right\rangle(H,J,T)$, and by some algebraic manipulations, we obtain
\begin{equation}
\left\langle m\right\rangle(H)=\frac{1}{2}H^{\frac{1}{3}},
\end{equation}
from which we get $\delta=3$. To find out $\gamma$ we solve

\begin{equation}
\frac{\partial^2 f}{\partial m\partial H}=-1+2(T-J)\frac{\partial f}{\partial m}+12m^2\frac{\partial f}{\partial m}=0,
\end{equation}
from which we get the magnetic susceptibility

\begin{equation}
\chi(T)=\frac{\partial f}{\partial m}=\frac{1}{2(T-J)-12m^2},
\end{equation}
where, by inserting $m(T)$ given in (\ref{mT}), we obtain

\begin{equation}
\chi(T)=\left\{\begin{array}{ll}
\frac{1}{2(T-J)}  & \hbox{if} \quad T\le T_c
\\
\frac{1}{10(T-J)}  & \hbox{if} \quad T\ge T_c
\end{array}\right.,
\end{equation}
from which $\gamma=1$. Summarizing, the critical exponents are those of a classical SBPT.

\subsection{Geometry and topology of the $\Sigma_{v,N}$'s and their link with the $\mathbb{Z}_2$-SBPT}
\label{rev_geom}

In Ref. \cite{bc} a topological sufficient condition for $\mathbb{Z}_2$-SB has been given (Theorem 1 in the paper). Simplifying the picture, Theorem 1 states that if the $\Sigma_{v,N}$'s are made at least by two connected components being one on the opposite side of the other with respect to the $m$-level set $\Sigma_{0,N}$ for $v\in [v',v'']$, then the $\mathbb{Z}_2$ symmetry is broken for the same $v$-values. We are assuming that the connected components are accessible to the representative point of the system. In this framework, the spontaneous magnetization $\langle m\rangle$ is the ensamble average of $m$ calculated on the connected component of the $\Sigma_{v,N}$ for which the density of states takes the global maximum. In the thermodynamic limit $v$ is selected by the temperature, i.e., $v=\left\langle v\right\rangle(T)$, then the spontaneous magnetization is in turn a $T$-function $\left\langle m\right\rangle(T)$. 

The $\Sigma_{v,N}$ of the model (\ref{Vrev}) is given by
\begin{equation}
\Sigma_{v,N}=\{\mathbf{q}\in M: m^4-Jm^2=v\},
\end{equation}
i.e.,
\begin{equation}
\Sigma_{v,N}=\cup_{m(v)\in I} \left(M\cap\Sigma_{m(v),N}\right),
\end{equation}
where $I$ is the set of the solutions of $-Jm^2+m^4=v$ given by 
\begin{equation}
m(v)=\pm\left(\frac{J\pm (J+4v)^{\frac{1}{2}}}{2}\right)^\frac{1}{2}.
\label{m(v)}
\end{equation}
We distinguish three cases:

\smallskip
(i) $v\in [-J/4, 0)$. Eq. (\ref{m(v)}) has four distinct solutions, each of them corresponds to a single connected component of the $\Sigma_{v,N}$ made by an $(N-1)$-ball of volume $e^{-(N-1)m(v)^2}$. These $\Sigma_{v,N}$'s satisfy the hypotheses of Theorem 1 in Ref. \cite{bc} which implies the $\mathbb{Z}_2$-SB for $T$-values such that $\left\langle v\right\rangle(T)\in [-J/4,0)$ (see Fig. \ref{fig_rev_sigmav}).

\smallskip
(ii) $v=0$. Eq. (\ref{m(v)}) has three distinct solutions one of which equals zero. The connected components of $\Sigma_{v,N}$ are three $(N-1)$-balls. In this case the $\mathbb{Z}_2$ symmetry is intact because the $(N-1)$-ball located at $m=0$ has a greater volume than the others. In the following we will see how to calculate it.

\smallskip
(iii) $v>0$. Eq. (\ref{m(v)}) has two distinct solutions. According to Theorem 1 in Ref. \cite{bc} the $\mathbb{Z}_2$ symmetry should be broken, but the $v$-values above zero are non-accessible to the representative point, so that $\Sigma_{0,N}$ plays the role of a critical $v$-level set separating the broken symmetry phase from the unbroken one.
\smallskip

The density of states at fixed $v$ is given by

\begin{eqnarray}
\omega_N(v)&=&\mu\left(\cup_{m(v)\in I} \left(M\cap\Sigma_{m(v),N}\right)\right)=\nonumber
\\
&=&\sum_{m(v)\in I}\mu\left(M\cap\Sigma_{m(v),N}\right)=\nonumber
\\
&=&\sum_{m(v)\in I}\int_{M\cap\Sigma_{m(v),N}}\frac{d\Sigma}{\Vert \nabla V\Vert}.
\label{rev_simgmav}
\end{eqnarray}
In order to evaluate the global maximum of $s_N(v,m)=1/N\log\omega_N(v,m)$, we need to know $\nabla V$ in (\ref{rev_simgmav}). It can be expressed in the coordinate system (\ref{cs}) as follows
\begin{equation}
\nabla V=\left(N(4m^3-2mJ),0,\cdots,0\right),
\end{equation}
whence $\|\nabla V\|=N|4m^3-2mJ|$. The last quantity is constant on the whole surface of $M\cap\Sigma_{m,N}$, so that it can be factorized in the integral (\ref{rev_simgmav}). In the limit $N\rightarrow\infty$ the contribution of $\|\nabla V\|$ to $\omega_N(m)^{1/N}$ is $1$, except at $m=0,\pm \sqrt{J/2}$ where it becomes infinite. Hence, we can replace the measure $\omega_N(m) $ with the standard volume $e^{-(N-1)m^2}$ defined in (\ref{volrev}). The singularities at $m=0,\pm \sqrt{J/2}$ do not cause any uncertainty in locating the spontaneous magnetization because of the structure of the $\Sigma_{v,N}$'s for $v=-J/4, 0$.

\begin{figure}
	\begin{center}
		\includegraphics[width=0.25\textwidth]{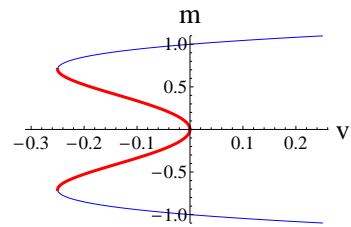}
		\caption{Solutions (\ref{m(v)}) for $J=1$. The red curve corresponds to the spontaneous magnetization as a function of the specific potential. The blue curves correspond to non-accessible regions of configuration space.}
		\label{fig_rev_mv}
	\end{center}
\end{figure}

The spontaneous magnetization as a $v$-function is plotted in Fig. \ref{fig_rev_mv}. 
At $v=0$ a topological change occurs. When $v$ reaches zero from below the two innermost $(N-1)$-balls of the $\Sigma_{v,N}$ joint becoming an $(N-1)$-ball alone. This is equivalent to what happens for a smooth potential when the $\Sigma_{v,N}$ crosses a critical level with a saddle point of index $1$. In this model the derivative along $m$ is negative and the derivatives along the $\widetilde{q}_i$'s, $i=1,\cdots,N-1$, are vanishing. Anyway, the shape of the $\Sigma_{v,N}$'s can be continuously deformed in such a way to make the last derivatives positive without changing the properties of the model. This is what happens in the model introduced in Sec. \ref{revs} equipped with a smooth potential. In this case a positive shift between the critical average thermodynamic potential and the critical topological level is entailed, in contrast to this model where they \emph{exactly} coincide.

\section{Revolution model with smooth potential}
\label{revs}

In this section we will modify the definition of the potential (\ref{Vrev}) in such a way to make it smooth. This is more realistic form a physical viewpoint. In that case we have constrained the configuration space $M$ into a sort of an $N$-dimensional 'spindle' such that $vol(M\cap \Sigma_{m,N})=e^{-(N-1)m^2}$. Here, we will follow a different way. We will attach at each point of the line passing through zero and orthogonal to the hyperplanes $\Sigma_{m,N}$'s a paraboloid weighted by the factor $e^{-m^2}$. In the coordinate system (\ref{cs}) the potential results to be
\begin{equation}
V=N(m^4-Jm^2)+\sum_{i=1}^{N-1}\left(\frac{\widetilde{q}_i}{e^{-m^2}}\right)^2.
\label{Vrevs}
\end{equation}
This potential has two global minima of value $-NJ/4$ whose coordinates are $\pm\left( \sqrt{J/2},0,\cdots,0\right)$ expressed in the system (\ref{cs}). As the model (\ref{Vrev}), this model has an $O(N-1)$ symmetry in the coordinates $\left(\widetilde{q}_1,\cdots,\widetilde{q}_{N-1}\right)$ beside the $\mathbb{Z}_2$.

\subsection{Canonical thermodynamic}

\begin{figure}
	\begin{center}
		\includegraphics[width=0.235\textwidth]{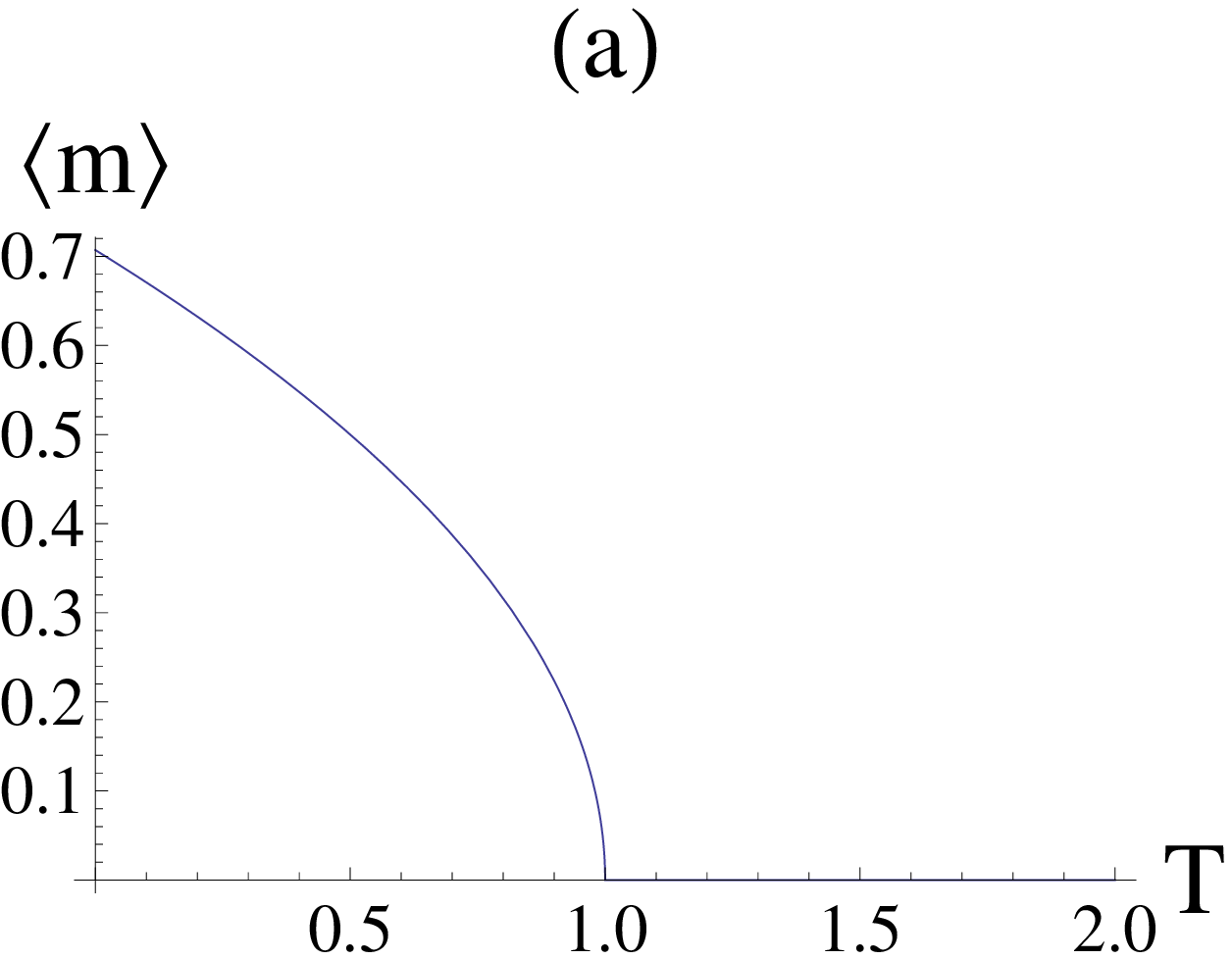}
		\includegraphics[width=0.235\textwidth]{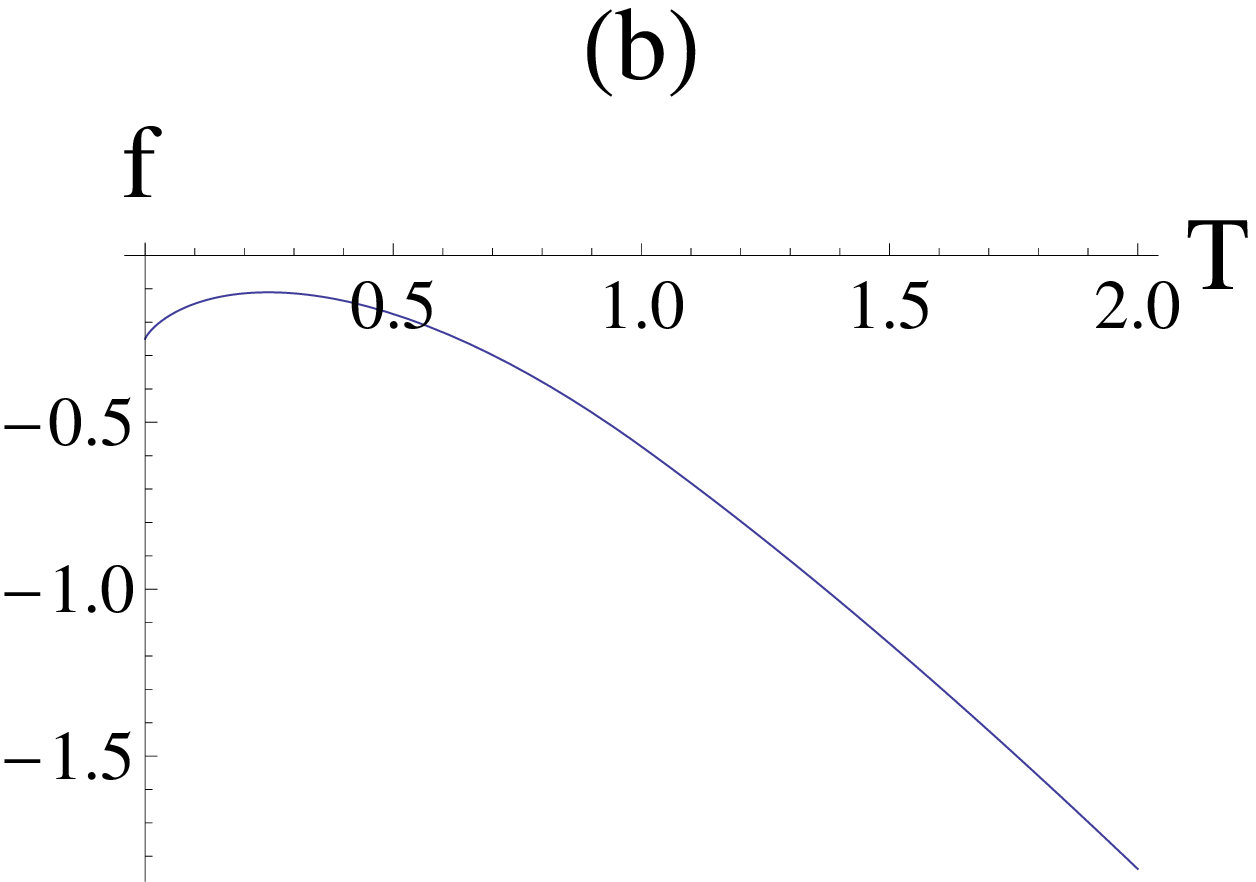}
		\includegraphics[width=0.235\textwidth]{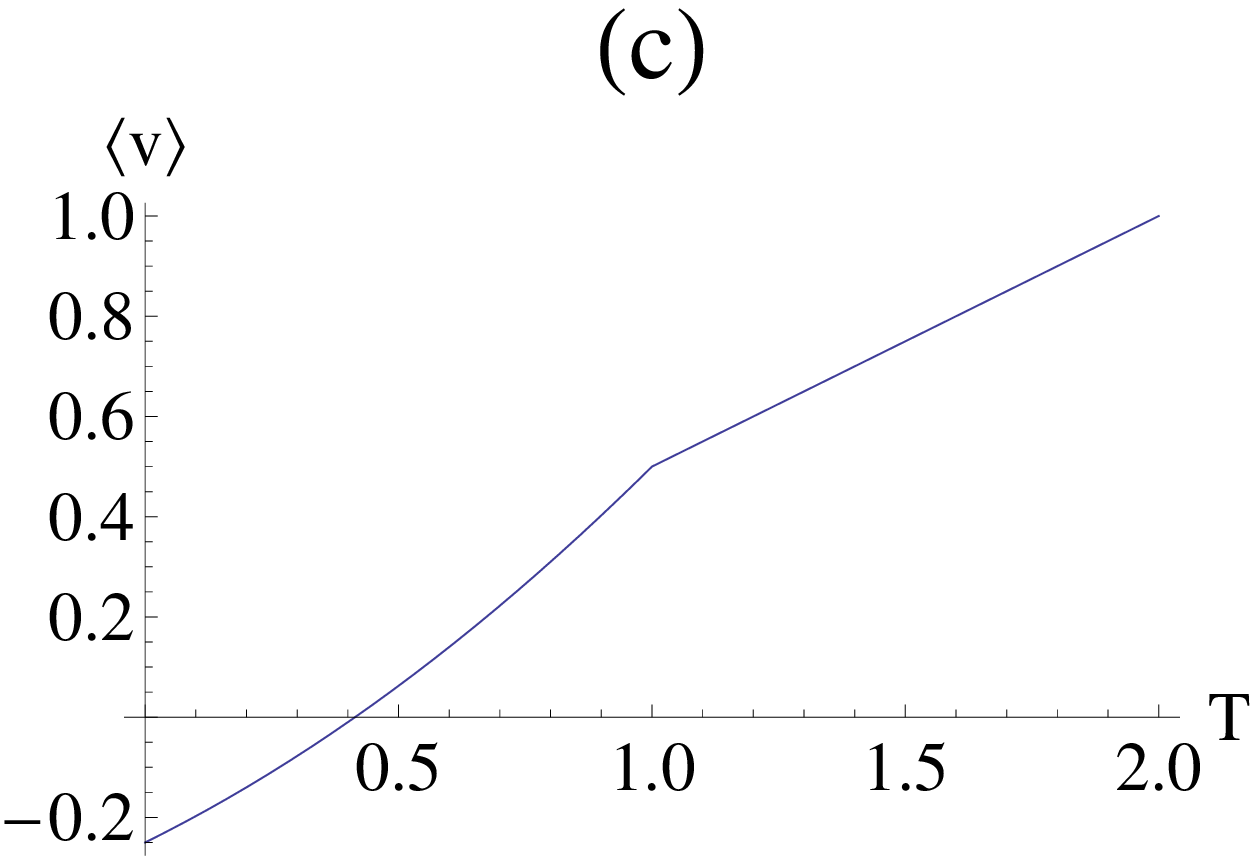}
		\includegraphics[width=0.235\textwidth]{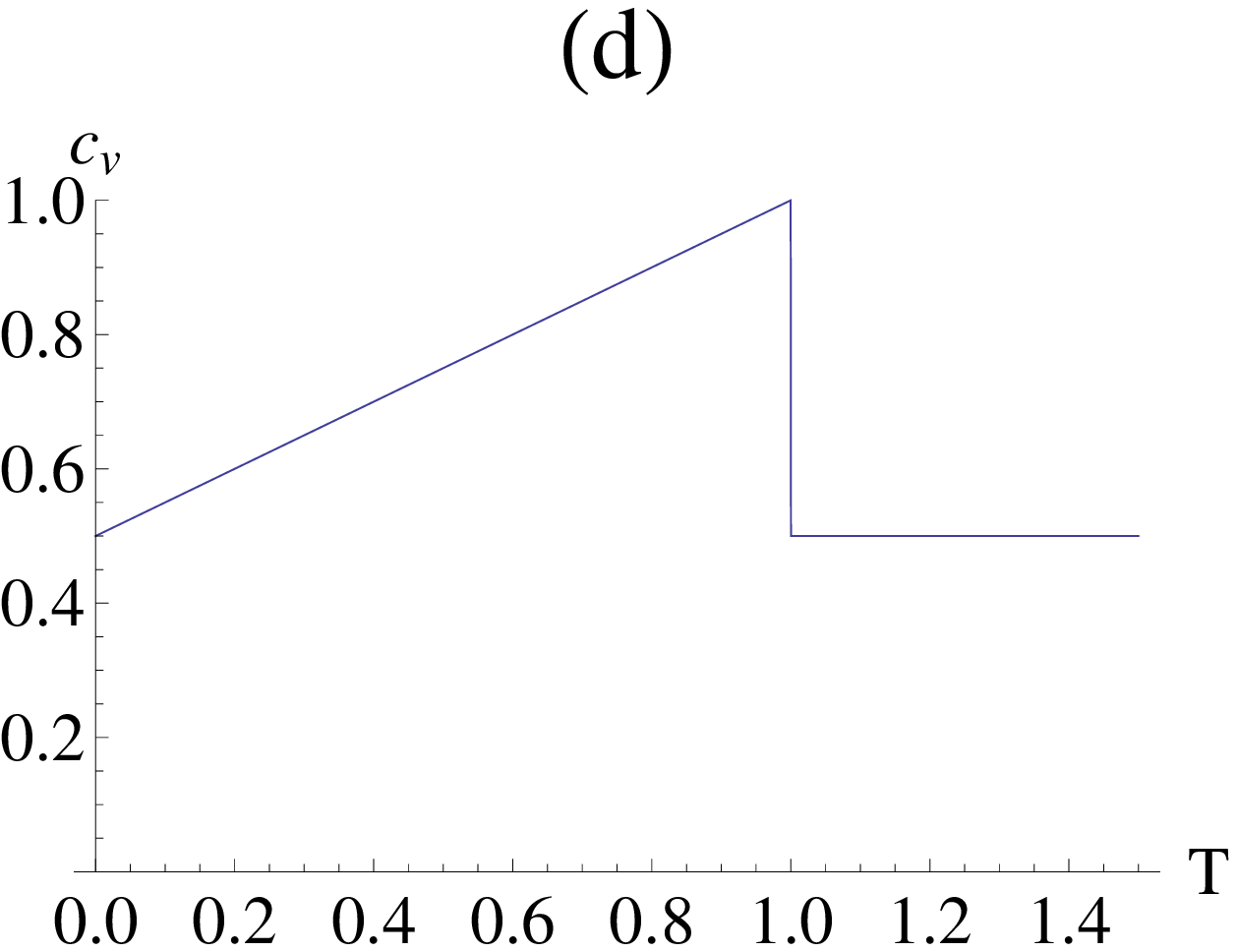}
		\caption{Model (\ref{Vrevs}) for $J=1$. (a), (b), (c), (d) Spontaneous magnetization, free energy, average potential, and specific heat versus the temperature, respectively.}
		\label{fig_revs}
	\end{center}
\end{figure}

The partition function is given by
\begin{equation}
Z_N=\sqrt{N}\int dm\,d\widetilde{\mathbf{q}}\, e^{-\frac{1}{T}\left(N(m^4-Jm^2)+e^{2m^2}\sum_{i=1}^{N-1}\widetilde{q}_i^2\right)},
\end{equation}
which can be re-written as
\begin{equation}
Z_N=\sqrt{N}\int dm\, e^{-\frac{N}{T}(m^4-Jm^2)}\left(\int dq\, e^{-\frac{e^{2m^2}}{T}q^2}\right)^{N-1}.
\end{equation}
By applying the Gaussian integral formula, in the large-$N$ limit, we get
\begin{eqnarray}
Z_N\simeq \sqrt{N}\int dm\, e^{-\frac{N}{T}\left(m^4+(T-J)m^2-\frac{T}{2}\ln\left(\pi T\right)\right)}.
\end{eqnarray}
In the thermodynamic limit, the free energy, the spontaneous magnetization, the average potential, and the specific heat are, respectively,
\begin{equation}
f=m^4(T-J)+m^2-\frac{T}{2}\ln(\pi T),
\end{equation}
\begin{equation}
\left\langle m\right\rangle=\left\{\begin{array}{ll}
\pm\left(\frac{1}{2}(J-T)\right)^{\frac{1}{2}}\quad &\hbox{if} \quad T\le T_c
\\
0\quad &\hbox{if} \quad T\ge T_c
\end{array} \right.,
\label{revs_mT}
\end{equation}
\begin{equation}
\left\langle v\right\rangle=\left\{\begin{array}{ll}
\frac{1}{2}T-\frac{1}{4}(J-T^2)\quad &\hbox{if} \quad T\le T_c
\\
\frac{1}{2}J\quad &\hbox{if} \quad T\ge T_c
\end{array} \right.,
\label{revs_vT}
\end{equation}
\begin{equation}
c_v=\left\{\begin{array}{ll}
\frac{1}{2}\left(1+T\right)\quad &\hbox{if} \quad T\le T_c
\\
0\quad &\hbox{if} \quad T\ge T_c
\end{array} \right.,
\end{equation}
where $T_c=J$ is the critical temperature (see Fig. \ref{fig_revs}). The SBPT is of the second order with classical critical exponents. 

\subsection{Dumbbell-shaped $\Sigma_{v.N}$'s at the origin of the $\mathbb{Z}_2$-SBPT}

In Ref. \cite{b_3} a new way of understanding a $\mathbb{Z}_2$-SBPT has been introduced. It is based on the concept of \emph{dumbbell-shaped} $\Sigma_{v,N}$'s defined in the following way. Each $\Sigma_{v,N}$ is in correspondence with the microcanonical density of states 
\begin{equation}
\omega_N(v,m)=\mu\left(\Sigma_{v,N}\cap \Sigma_{m,N}\right)=\int_{\Sigma_{v,N}\cap \Sigma_{m,N}}\frac{d\Sigma}{\|\nabla V \wedge\nabla M\|}.
\label{omega_gram}
\end{equation}
A $\Sigma_{v,N}$ is defined to be dumbbell-shaped if the microcanonical entropy $s_N(v,m)$ does not take the global maximum at $m=0$ at fixed $v$. An equivalent definition can be given in terms of $\omega_N(v,m)^{1/N}=e^{s_N}$. Note that this definition is valid for each $N$, so that the study of SBPTs based on this framework is suitable, non only in the thermodynamic limit, but also in the case of finite $N$.

The main result of this approach is summarized in a straightforward theorem stating that the $\mathbb{Z}_2$ symmetry is broken if, and only if, the $\Sigma_{\left\langle v\right\rangle(T),N}$ corresponding to the temperature $T$ is dumbbell-shaped for all $N>N_0$, where $N_0$ is a fixed natural.
Furthermore, the critical average potential $v_c=\left\langle v\right\rangle(T_c)$ is \emph{exactly} in correspondence with the $\Sigma_{v,N}$ which is the boundary between the dumbbell-shaped $\Sigma_{v,N}$'s and those which are not. This $\Sigma_{v,N}$ is called critical.

\begin{figure}
	\begin{center}
		\includegraphics[width=0.235\textwidth]{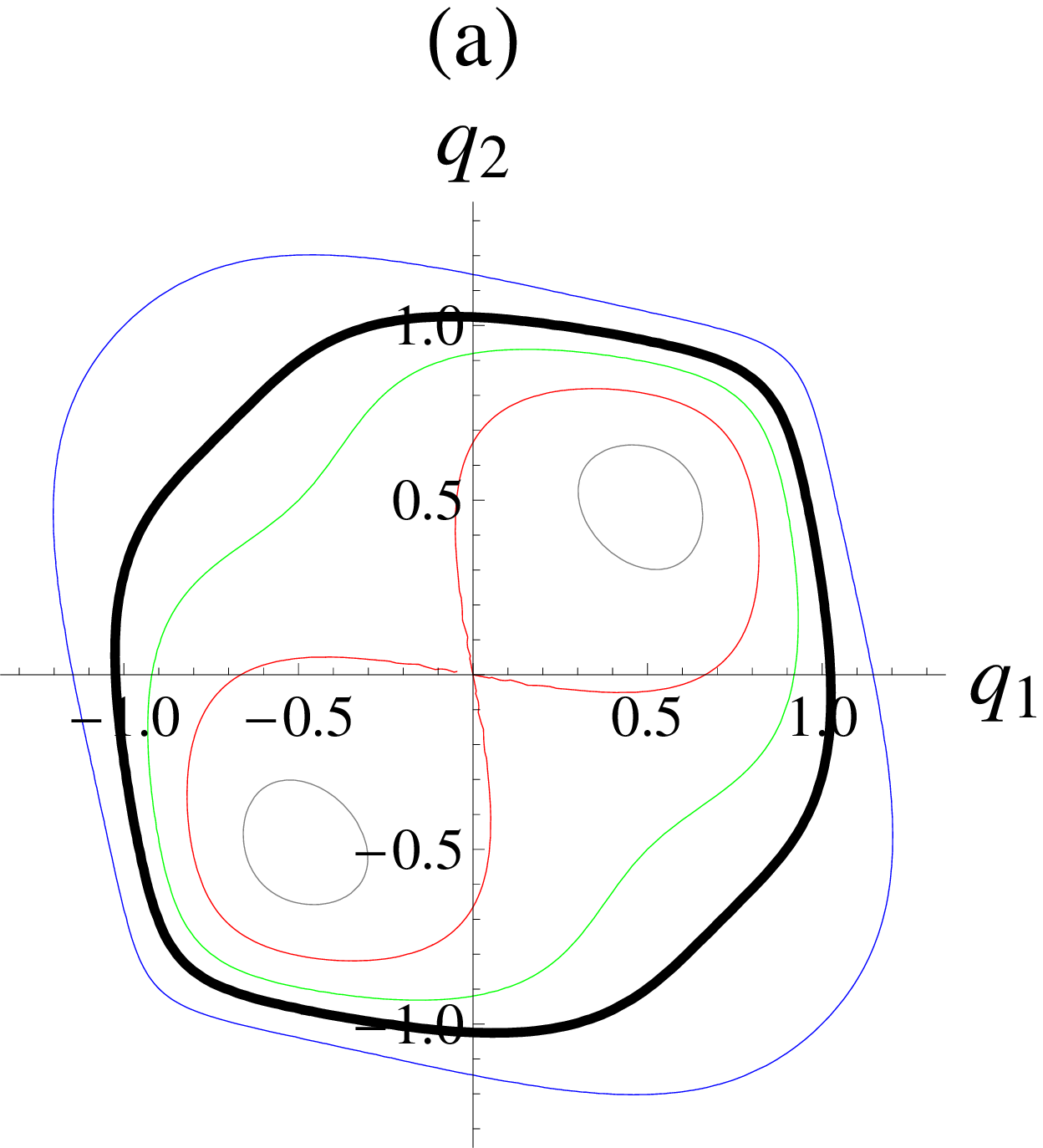}
		\includegraphics[width=0.235\textwidth]{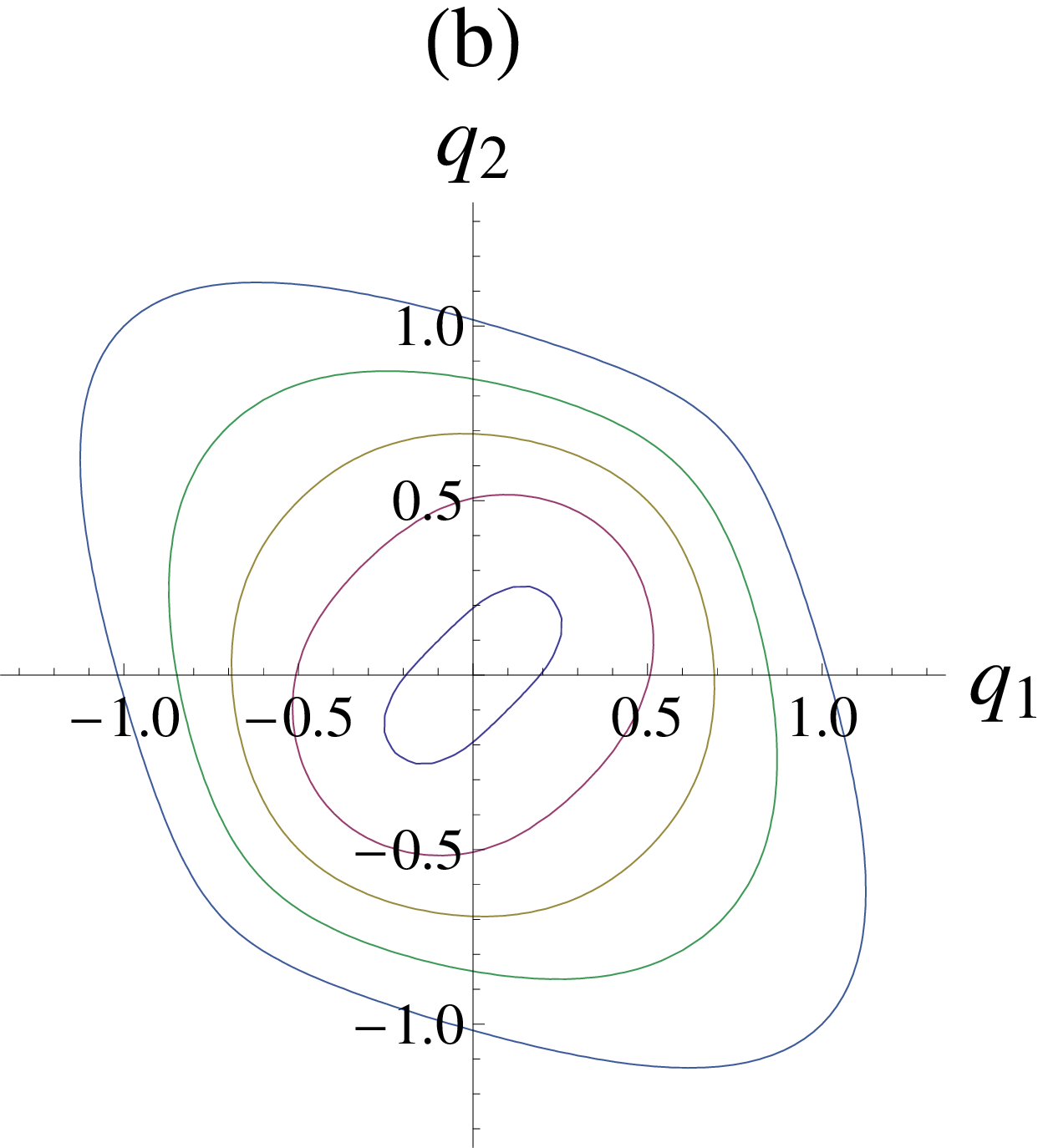}
		\caption{(a) Some $\Sigma_{v,N}$'s for $N=2$ of the model (\ref{Vrevs}) for $J=1$ and $v=-0.2, ~0, ~0.25 ,~0.5,~ 1$. $\Sigma_{0.5,2}$ is marked. It is the boundary between the dumbbell-shaped $\Sigma_{v,N}$'s for $v\in [-0.25,0.5]$ and those which are not for $v\geq 0.5$. (b) The same of the panel (a) for $J=0$ and $v=0.01, ~0.1, ~0.25 ,~0.5,~ 1$.}
		\label{fig_revs_sigmav}
	\end{center}
\end{figure}

The topology of the $\Sigma_{v,N}$'s can be discovered by Morse theory. The key concept is the attachment of an $i$-handle at each critical point of index $i$. The index is defined as the number of negative eigenvalues of the Hessian matrix. The potential (\ref{Vrevs}) has two critical levels. A critical level is a $v$-level set which at least contains a critical point. The lower one contains two critical points: $\pm\left(\sqrt{J},\cdots,\sqrt{J}\right)$. The Hessian matrix is diagonal $H_V=2\,diag\left(5NJ,e^{2m^2},\cdots,e^{2m^2}\right)$, thus the indexes are $0$. This corresponds to the attachment of two $0$-handles, so that the topology is that of the union of two disjointed $N$-spheres. The other critical level contains the saddle point $(0,\cdots,0)$, for which $H_V=2\,diag\left(-NJ,e^{2m^2},\cdots,e^{2m^2}\right)$, thus the index is $1$. After attaching an $1$-handle, the topology becomes that of a single $N$-sphere. Summarizing,
\begin{equation}
\Sigma_{v,N}\sim\left\{\begin{array}{ll}
\mathbb{S}^{N-1}\quad &\hbox{if} \quad v>0
\\
\hbox{critical} &\hbox{if} \quad v=0
\\
\mathbb{S}^{N-1}\cup \mathbb{S}^{N-1} &\hbox{if} \quad 0>v\geq-\frac{J}{4}
\\
\emptyset &\hbox{if} \quad v<-\frac{J}{4}
\end{array}\right.,
\end{equation}
where '$\sim$' stands for 'is homeomorphic to' (see Fig. \ref{fig_revs_sigmav}). 
There exists only a topological change at $v=0$. This potential satisfy the hypotheses of Theorem 1 in Ref. \cite{bc} for $v\in [-J/4,0)$, so that the $\mathbb{Z}_2$-SB is guaranteed for $T\in [0,T')$, where $T'=\left\langle v\right\rangle^{-1}(T)=-1+\sqrt{1+J}$ for $\left\langle v\right\rangle=0$, by topological reasons. Indeed, the $\Sigma_{v,N}$'s are made up by two connected components which are located on the opposite side of the $m$-level set $\Sigma_{0,N}$ with respect to the other. The critical average potential $\langle v\rangle_c=J/2$ is located above the unique critical $v$-level set $\Sigma_{0,N}$. This is due to the presence of dumbbell-shaped $\Sigma_{v,N}$'s in the interval $[0,J/2)$. Indeed, they imply the $\mathbb{Z}_2$-SB according to the theorem in Ref. \cite{b_3}. 

\begin{figure}
	\begin{center}
		\includegraphics[width=0.235\textwidth]{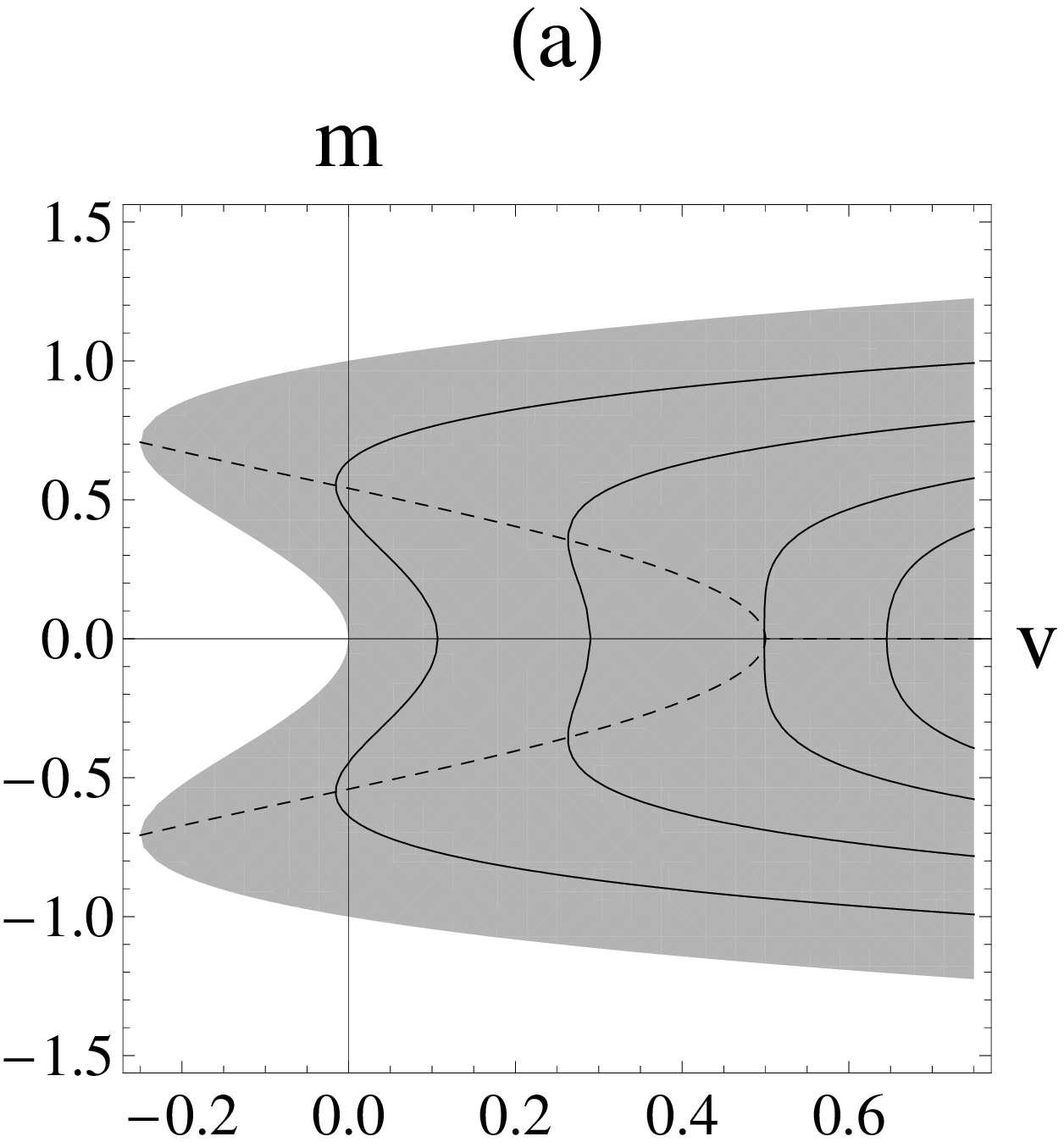}
		\includegraphics[width=0.235\textwidth]{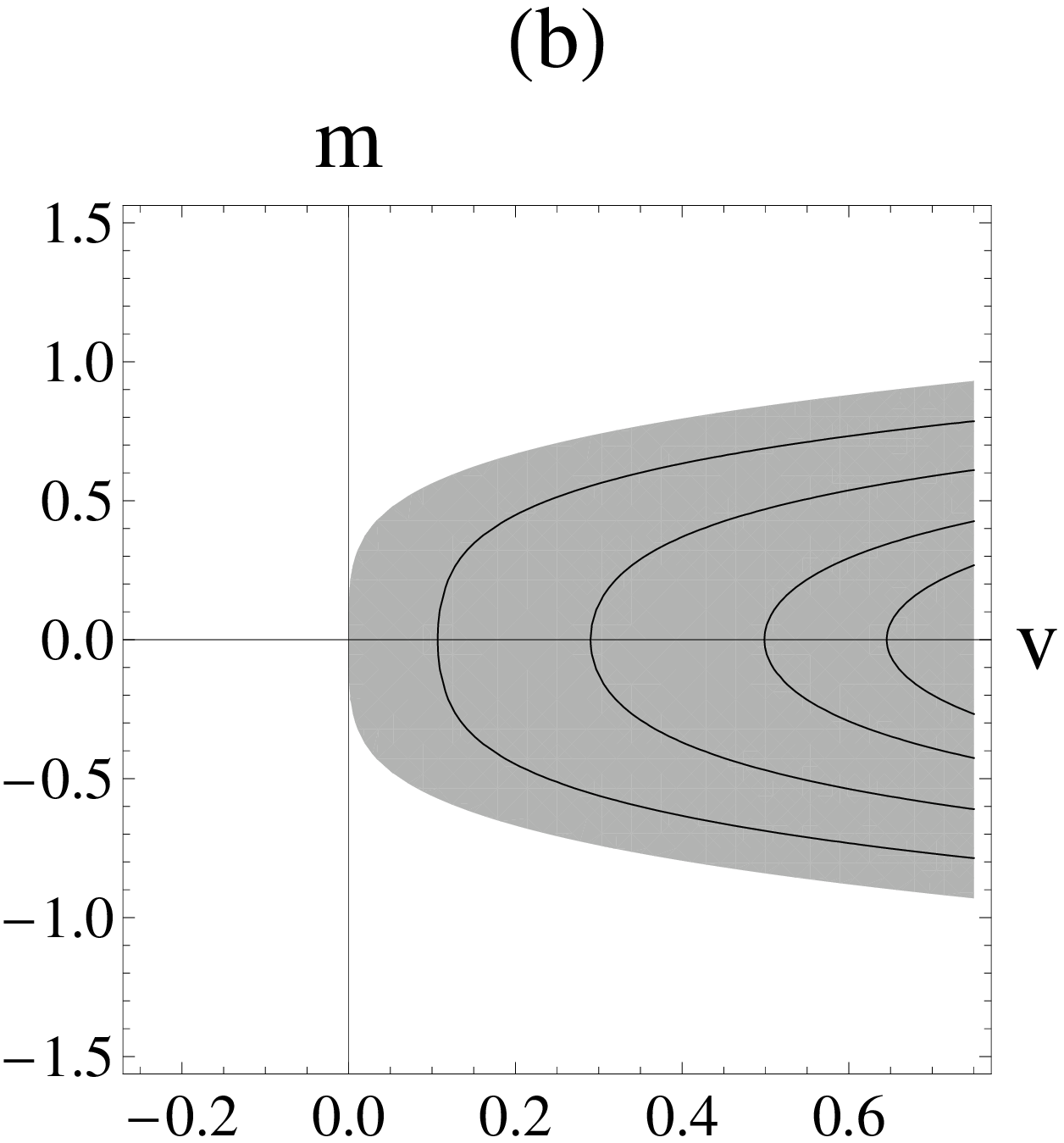}
		\caption{(a) Model (\ref{Vrevs}) for $J=1$. Contour plot of the microcanonical entropy $s(v,m)$ (\ref{revs_a}), the dark region surrounded by the curve of equation $v=-m^2+m^4$ is the domain. $v=\langle v\rangle_c=0.5$ is the boundary between the dumbbell-shaped $\Sigma_{v,N}$'s from the non-dumbbell-shaped ones. (b) The same of the panel (a) for $J=0$.}
		\label{fig_revs_s}
	\end{center}
\end{figure}

\begin{figure}
	\begin{center}
		\includegraphics[width=0.35\textwidth]{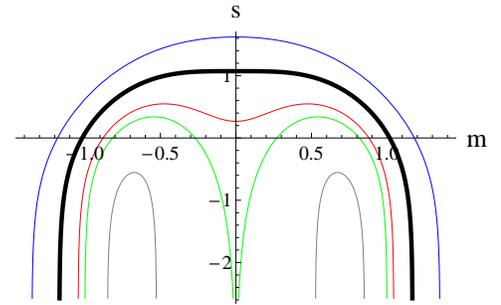}
		\caption{Microcanonical entropy (\ref{revs_a}) of the model (\ref{Vrevs}) for $J=1$ and $v=-0.2, ~0, ~0.1 ,~0.25,~ 1.5$. The marked curve is the boundary between the dumbbell-shaped $\Sigma_{v,N}$'s for $v\in [-0.25,0.5]$ and those which are not for $v\geq 0.5$.}
		\label{fig_revs_sm}
	\end{center}
\end{figure}

The simplicity of this model, in particular the presence of the $O(N-1)$ symmetry, allows to identify the dumbbell-shaped $\Sigma_{v,N}$'s by the explicit calculation of the density of states $\omega_N(v,m)$. Indeed, $\Sigma_{v,N}\cap\Sigma_{m,N}$ is an $(N-1)$-sphere defined by the following implicit equation
\begin{equation}
Nv=N(m^4-Jm^2)+e^{2m^2}\sum_{i=1}^{N-1}\widetilde{q}_i^2.
\end{equation}
The radius $R$ is given by
\begin{equation}
R^2=\sum_{i=1}^{N-1}\widetilde{q}_i^2=Ne^{-2m^2}(v-m^4+Jm^2),
\end{equation}
and the volume is given by
\begin{equation}
vol\left(\Sigma_{v,N}\cap\Sigma_{m,N}\right)=\frac{2\pi^{\frac{N-1}{2}}}{\Gamma\left(\frac{N-1}{2}\right)}R^{N-2}.
\end{equation}
To calculate $\omega_N(v,m)$ we need to take into account the Gramian square root $\Vert\nabla V\wedge\nabla M\Vert$ in (\ref{omega_gram}). In the coordinate system (\ref{cs})
\begin{equation}
\begin{split}
\nabla V=\left(N(4m^3-2Jm),2e^{2m^2}\widetilde{q}_1,\cdots,2e^{2m^2}\widetilde{q}_{N-1}\right),
\end{split}
\end{equation}
and
\begin{equation}
\nabla M=\left(N,0,\cdots,0\right).
\end{equation}
Thus,
\begin{equation}
\Vert\nabla V\wedge\nabla M\Vert^2=\det\left(
\begin{matrix}
\nabla V\cdot\nabla V & \nabla M\cdot\nabla V
\\
\nabla V\cdot\nabla M & \nabla M\cdot\nabla M
\end{matrix}
\right),
\end{equation}
from which, after some trivial algebraic manipulations, we get
\begin{equation}
\Vert\nabla V\wedge\nabla M\Vert=2 Ne^{2m^2}R.
\end{equation}
The last term is constant onto the whole integration support of the integral (\ref{omega_gram}), so that it can pass under the integral sign. As $N\to\infty$ we find out the entropy
\begin{eqnarray}
& &s(v,m)=\lim_{N\rightarrow\infty}\log\omega_N(v,m)^{\frac{1}{N}}=\nonumber
\\
& &=-m^2+\frac{1}{2}\log(v-m^4+Jm^2)+\frac{1}{2}\log(2\pi e).
\label{revs_a}
\end{eqnarray}
See Fig. \ref{fig_revs_s} and \ref{fig_revs_sm} for plots.

\begin{figure}
	\begin{center}
		\includegraphics[width=0.235\textwidth]{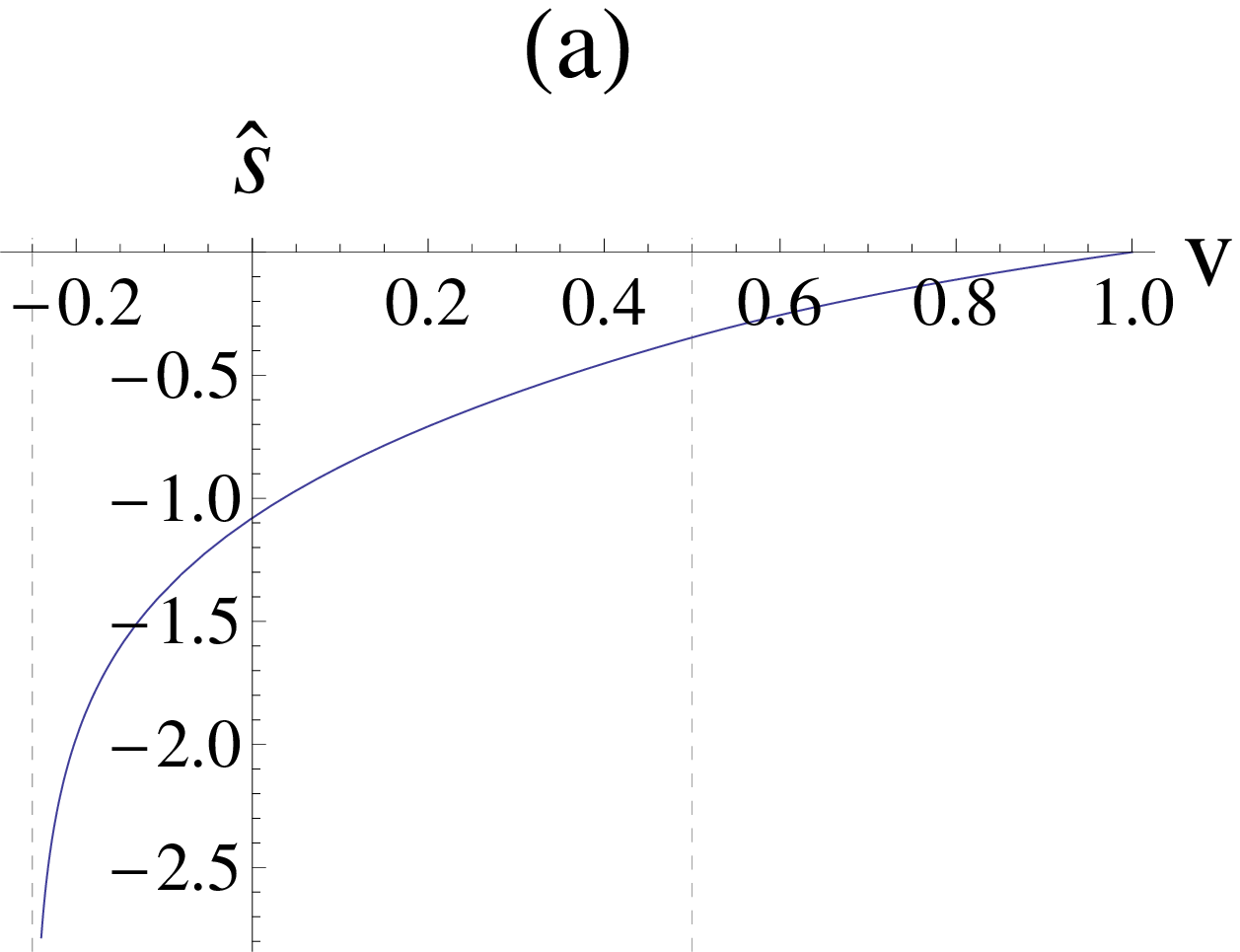}
		\includegraphics[width=0.235\textwidth]{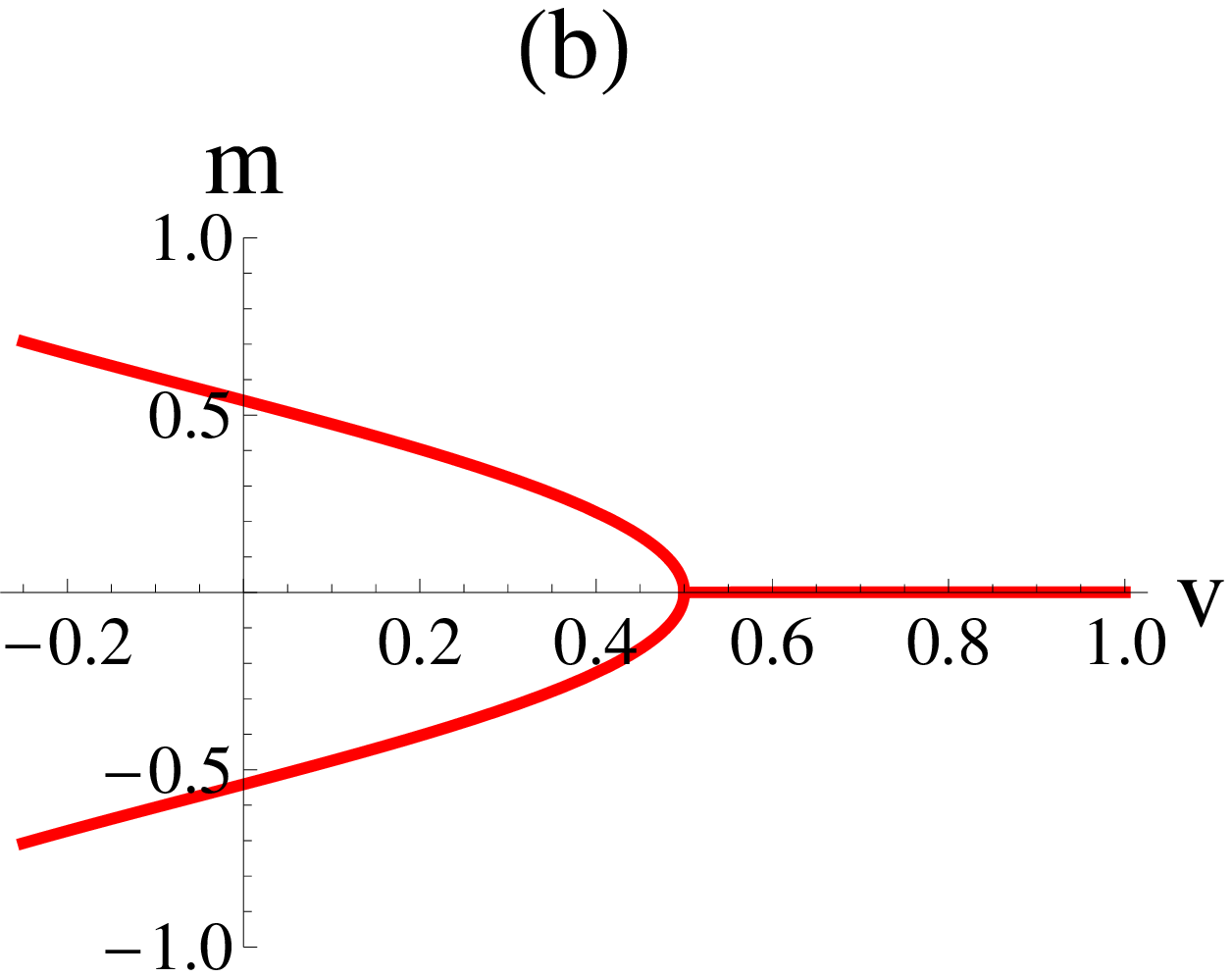}
		\caption{(a) Microcanonical entropy of the model (\ref{Vrevs}) for $J=1$ as a function of the specific potential. At $v=\langle v\rangle_c=0.5$ there is a third-order singularity. (b) Spontaneous magnetization as a function of the specific potential. The SBPT is at the same $v$-value.}
		\label{fig_revs_sv_mv}
	\end{center}
\end{figure}

According to the definition given in Ref. \cite{b_3}, a $\Sigma_{v,N}$ is called dumbbell-shaped if the related $s(v,m)$ does not take the global maximum at $m=0$. For $v\in[-J/4,0)$ the $\Sigma_{v,N}$'s are dumbbell-shaped because they are the union of two connected components (see Fig. \ref{fig_revs_sigmav}). The solution with respect to $v$ of the following equation
\begin{equation}
\frac{\partial s(v,m)}{\partial m}=0
\end{equation}
gives the spontaneous magnetization as a $v$-function 
\begin{equation}
m(v)=\begin{cases}
\pm\left(1-\left(v+\frac{1}{2}\right)^\frac{1}{2}\right)^\frac{1}{2} & \text{ if }\quad -\frac{1}{4}\le v\le \frac{1}{2}
\\
0 & \text{ if } \quad v\ge\frac{1}{2}
\end{cases},
\end{equation}
(see Fig. \ref{fig_revs_sv_mv}). 

By inserting eq. (\ref{revs_vT}) in the last one we get eq. (\ref{revs_mT}). It is a remarkable fact that we could get the same result in two independent ways: via canonical thermodynamic and via geometry of the potential landscape. Consider $v\geq 0$. To discover weather a $\Sigma_{v,N}$ is dumbbell-shaped is sufficient to set to zero the second partial derivative of $s(v,m)$ with respect to $m$ at $m=0$
\begin{equation}
\left.\frac{\partial^2 s(v,m)}{\partial m^2}\right|_{m=0}=2 v-J=0.
\end{equation}
Therefore, $v=J/2$ is the boundary between the dumbbell-shaped $\Sigma_{v,N}$'s from those which are not. In particular, the $\Sigma_{v,N}$'s are dumbbell-shaped for $v<J/2$ (see Fig. \ref{fig_revs_sigmav}). $\Sigma_{J/2,N}$ plays the role of a critical $v$-level set.

From the thermodynamic viewpoint, we know that the critical average potential is just $\left\langle v\right\rangle_c=J/2$, so that the canonical thermodynamic picture of the $\mathbb{Z}_2$-SBPT is in perfect agreement with the geometric picture based on the of dumbbell-shaped $\Sigma_{v,N}$'s framework. Furthermore, we note that the microcanonical entropy $\hat{s}(v)$ can be obtained by a maximization process of $s(v,m)$ with respect to $m$ as it has been made in Ref. \cite{hk} for the mean-field $\phi^4$ model (see Fig. \ref{fig_revs_sv_mv})
\begin{equation}
\hat{s}(v)=\begin{cases}e^{\left(v+\frac{1}{2}\right)^\frac{1}{2}-1} \left(\left(v+\frac{1}{2}\right)^\frac{1}{2}-\frac{1}{2}\right)^\frac{1}{2} & \text{if}~~ -\frac{1}{4}\le v\le\frac{1}{2}
\\
e^\frac{1}{2} & \text{if}~~~v\ge\frac{1}{2}
\end{cases},
\end{equation}
and $s(T)$ can be calculated by inserting eq. (\ref{revs_vT}) in the equation above getting
\begin{equation}
s(T)=\begin{cases}\frac{1}{2} \log \left(\frac{1}{4} \left(T^2+2 T-1\right)\right) & \text{if}~~ 0\le T\le 1
\\
\frac{1}{2} \left(T+\log \left(\frac{T}{2}\right)-1\right) & \text{if}~~~T\ge 1
\end{cases}.
\end{equation}
The same for $m(T)$.

\subsection{The case at finite N}
\label{revsN}

 The formula of the microcanonical entropy for finite $N$ is the following
\begin{equation}
\begin{split}
s_N(v,m)=-\frac{N-5}{N}m^2+
\frac{N-5}{2N}\log N+\frac{N-1}{2N}\log\pi+\\
\frac{N-3}{2N}\log\left(v-m^4+Jm^2\right)-\frac{1}{N}\log\Gamma\left(\frac{N-1}{2}\right).
\end{split}
\label{revs_sN}
\end{equation}
There are no substantial differences in the shape of the graph compared to the case $N=\infty$ discussed in previous section. The definition of SBPT given in Ref. \cite{b_3} holds also for finite $N$. Therefore, the model (\ref{Vrevs}) undergoes the spontaneous symmetry breaking of its $\mathbb{Z}_2$ symmetry for each $N>5$. The smaller $N$'s must be excluded because the formula (\ref{revs_sN}) makes no sense. The critical potential $v_{c,N}$ turns out to be an $N$-function tending to $v_c=1/2$ for $N\to\infty$
\begin{equation}
v_{c,N}=\frac{J (N-3)}{2 (N-5)}.
\end{equation}

We do not enter into the discussion of what the temperature for finite $N$ may be, in particular the critical temperature, because this is not the right place to deal with this problem. We merely observe that we do not see any substantial difference with respect to the thermodynamic limit except in the fact that the fluctuations of the physical quantities are non-vanishing.

\subsection{On the origin of the PT}

In Refs. \cite{ccp2,nc,km,mhk,nkmc,gfp,gfp1,gfp2,cccp,ccp,ccp1,ck,ck1,fp,fp1,ks,kss1,kss,hk,pettini} a great effort has been put in trying to understand the deep origin of a PT meant as a loss of analyticity in the thermodynamic functions. Here, our purpose is to make some considerations about that question. 

The free energy $f$ in the $(m,T)$-plane is an analytic function in the thermodynamic limit, but, e.g., this is not the case of the spontaneous magnetization as a $T$-function. By resorting to Dini theorem (or the implicit function theorem) we know that the graph of the zeroes of the partial derivative with respect to $m$ of $f$, i.e., the spontaneous magnetization, is an analytic function too. More precisely, if $f(m,T)\in C^k$, then also $m(T)\in C^k$ for $k=1,\cdots,\infty$. 

The singular point in the graph of $m(T)$ arises because it is the union of two analytic branches connected by a non-analytic point. The two branches are the line $m(T)=0$ for $T\ge T_c$, and the parabola $m(T)=\pm (1/2(T_c-T))^{1/2}$ for $T\leq T_c$, which touch each other at $(0,T_c)$. In Ref. \cite{hk} it has been shown that a non-analyticity in the microcanonical entropy $s(v)$ stems from a maximization process of the entropy $s(v,m)$ with respect to $m$, which is strictly correlated to what aforementioned. This holds only if the graph of $s(v,m)$ is non-concave. There is no way to generate such a non-analyticity starting from a strictly concave graph. 

Generally, the presence of a singularity in the thermodynamic functions is associated with the presence of spontaneous SB, but there exist cases in which this is not true. For example, in the hypercubic model introduced in Ref. \cite{bc}, the first-order PT is not related to the $\mathbb{Z}_2$-SB, but rather it stems from the fact that the potential is not a continuous function of coordinates. Indeed, it assumes two discrete values only. To conclude, at this stage we cannot suggest any unified origin for a PT to occur.

\section{Other critical exponents and universality classes}
\label{rev_oce}

We can generalize the definition of the potential (\ref{Vrevs}) as follows
\begin{equation}
V=N\left(m^{2l}-Jm^{2k}\right)+\sum_{i=1}^{N-1}\left(\frac{\widetilde{q}_i}{e^{-m^{2k}}}\right)^2,
\label{Vrevs2d}
\end{equation}
with $k,l$ naturals such that $0<k<l$. The same can be made for the potential (\ref{Vrev}) by redefined the volume of the subsets of configuration space $M$ at constant $m$ (\ref{volrev}) as $e^{-(N-1)m^{2k}}$. For suitable chooses of $k,l$ the revolution model belongs also to further universality classes than the classical one. 

\begin{figure}
	\begin{center}
		\includegraphics[width=0.35\textwidth]{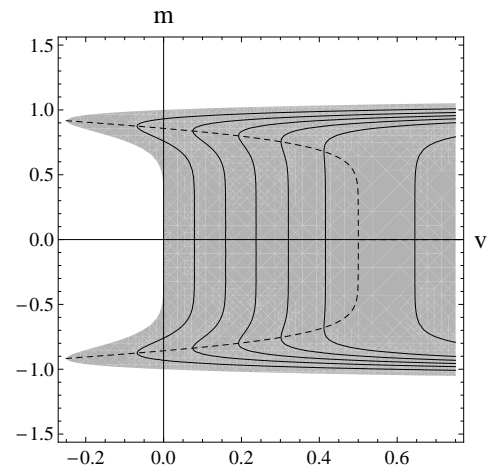}
		\caption{Model (\ref{Vrevs2d}) for $J=1$. Contour plot of the microcanonical entropy $s(v,m)$ (\ref{fig_Vrevs2d_svm}), the dark region surrounded by the curve of equation $v=-m^8+m^{16}$ is the domain. The dashed curve is the spontaneous magnetization.}
		\label{fig_2d_svm}
	\end{center}
\end{figure}

For example, we will calculate the critical exponents for the case $k=8$ and $l=16$ corresponding to the universality class of the short-range 2d Ising model. The free energy in the thermodynamic limit is 
\begin{equation}
f=m^{16}+(T-J)m^8-\frac{T}{2}\log(\pi T),
\end{equation}
and the critical temperature is $T_c=J$.
By solving $\partial f/\partial m=0$ we obtain the spontaneous magnetization
\begin{equation}
\left\langle m\right\rangle=\begin{cases}
\pm\left(\frac{1}{2}(J-T)\right)^{\frac{1}{8}} & \text{ if }\,\, T\leq T_c
\\
0 & \text{ if } \,\,T\geq T_c
\end{cases},
\label{Vrevs2d_mT}
\end{equation}
where $T_c=J$, whence $\beta=1/8$. $\left\langle v\right\rangle(T)$ and $c_v(T)$ are the same of the model (\ref{Vrevs}), so that $\alpha=0$. After inserting the external magnetic field $H$, the free energy becomes
\begin{equation}
f=m^{16}+(T-J)m^8-mH-\frac{T}{2}\log(\pi T),
\end{equation}
from which, by following the same procedure of Sec. \ref{rev_emf}, we get
\begin{equation}
\left\langle m\right\rangle(H)\propto H^{\frac{1}{15}},                 
\end{equation}
whence $\delta=15$, and
\begin{equation}
\chi(T)\propto |T_c-T|^{-\frac{7}{4}},
\end{equation}
whence $\gamma=7/4$, as promised. We can also give the microcanonical entropy (see Fig. \ref{fig_2d_svm} for a plot)
\begin{eqnarray}
s(v,m)&=&\lim_{N\rightarrow\infty}\log\omega_N(v,m)^{\frac{1}{N}}=\nonumber
\\
&=&-m^8+\frac{1}{2}\log(v-m^{16}+Jm^8)+\nonumber
\\
&+&\frac{1}{2}\log(2\pi e).
\label{fig_Vrevs2d_svm}
\end{eqnarray}

However, there is a difference that it is worth be noted with respect to the short-range 2d Ising model. In the case of the latter the specific heat undergoes a logarithmic  divergence at the SBPT. We are currently unable to justify this difference. We merely suggest that it may be related to the fact that the revolution model with potential (\ref{Vrevs2d}) belongs to the class of long range systems unlike the short-range 2d Ising model.

\section{Connection with physical models}
\label{phi4}

In this section we will deal with a physical model with the aim of highlighting the connection with the revolution model with smooth potential (\ref{Vrevs}) proposed in this article.

\subsection{The mean-field $\phi^4$ model}
\label{phi4mf}

We recall that the potential of the mean-field $\phi^4$ model is as follows
\begin{equation}
V=\sum^{N}_{i=1}\left(\frac{\phi_i^4}{4}-\frac{\phi_i^2}{2}\right)-\frac{J}{2N}\left(\sum^{N}_{i=1}\phi_i\right)^2.
\end{equation}
The model is known to undergo a second-order $\mathbb{Z}_2$-SBPT with classical critical exponents. In Ref. \cite{hk} the authors have been able to calculate the thermodynamic limit of the microcanonical entropy $s(v,m)$ by the large deviations theory. The microcanonical entropy $\hat{s}(v)$ is obtained by a process of maximization of $s(v,m)$ with respect to $m$
\begin{equation}
\hat{s}(v)=\max_{m}s(v,m).
\label{Vphi4}
\end{equation}
The domain of $s(v,m)$ is a non-convex subset of the plane $(v,m)$, and $s(v,m)$ is a non-concave function. The critical average potential $\langle v\rangle_c$ of the $\mathbb{Z}_2$-SBPT is located in such a way to divide the concave sections of the graphs $s(v,m)$ at fixed $v$ for $v\ge \langle v\rangle_c$ from the non-concave ones for $v<\langle v\rangle_c$. The graph of $s(v,m)$ (Fig. 5 in Ref. \cite{k} and Fig. 2 in Ref. \cite{hk}) is qualitatively \emph{identical} to that of the model (\ref{Vrevs}) reported in Fig. \ref{fig_revs_s}. This holds both for $J>0$ and $J=0$.

In Refs. \cite{aarz,b_1} the topology of the $\Sigma_{v,N}$'s has been studied exhaustively by means of Morse theory. The following three cases have been delineated:

\smallskip
(i) $v\in [v_{min},v_t)$. $v_{min}=-(1+J)^2/4$ is the global minimum of the potential. $v_t$ depends on the coupling constant $J$, and $v_t<-1/4$. The $\Sigma_{v,N}$'s are homeomorphic to the union of two disjointed $N$-spheres. The thermodynamic critical potential $\langle v\rangle_c$ of the $\mathbb{Z}_2$-SBPT can be less than $0$, but $\langle v\rangle_c>v_t$ holds for all $J>0$. 

\smallskip
(ii) $v\in [v_t,0]$. There is a huge amount of critical points growing as $e^N$ as a consequence of the topological changes. We can say that the whole interval $[v_t,0]$ plays the role of a critical level, because it discriminates between the $\Sigma_{v,N}$'s homeomorphic to two disjointed $N$-spheres from those homeomorphic to an $N$-sphere alone. In Ref. \cite{b_5} it has been shown how to reduce this critical interval to a single $v$-value corresponding to a critical $v$-level set containing a single critical point. Furthermore, as $J\rightarrow+\infty$, $v_t\rightarrow-1/4^-$.

\smallskip
(iii) $v\in (0,+\infty)$. The $\Sigma_{v,N}$'s are homeomorphic to an $N$-sphere. 

\smallskip
Let us try to interpret this picture in the framework of the dumbbell-shaped $\Sigma_{v,N}$'s. In the case (i) the hypotheses of Theorem 1 in Ref. \cite{bc} are satisfied, thus the topology of the $\Sigma_{v,N}$'s implies the $\mathbb{Z}_2$-SB. This is in accordance with $\langle v\rangle_c>v_t$ for all $J>0$ because the spontaneous magnetization cannot vanish below $v_t$. Since Theorem 1 in Ref. \cite{bc} is a special case of the theorem given in Ref. \cite{b_3}, also the hypotheses of the last one are satisfied. In the case (ii) the hypotheses of Theorem 1 in \cite{bc} are not satisfied, so that only the theorem in Ref. \cite{b_3} may implies the $\mathbb{Z}_2$-SB. Indeed, the $\Sigma_{v,N}$'s may be dumbbell-shaped below $\langle v\rangle_c$ and non-dumbbell-shaped above $\langle v\rangle_c$ (if $\langle v\rangle_c<0$) independently on their intricate topology. Finally, the same of the case (ii) holds for the case (iii). The difference is that in the last case the $\Sigma_{v,N}$'s are all diffeomorphic to an $N$-sphere alone. However, this difference is not significant.

$\Sigma_{\langle v\rangle_c,N}$ plays the role of the critical level in the sense of the theorem in Ref. \cite{b_3} because it separates the dumbbell-shaped $\Sigma_{v,N}$'s from those which are not. For the sake of precision, we aspect that at fixed $N$ the critical $\Sigma_{v,N}$ in the above-specified sense is not located exactly at $\langle v\rangle_c$, but that there exists a sequence of critical $\Sigma_{v_N,N}$ such that $v_N\rightarrow \langle v\rangle_c$ for $N\rightarrow\infty$. Further analytic and numerical studies may check this conjecture.

The potential of the mean-field $\phi^4$ model is made by a mean-field-like interacting part with the addition of a constraint given by the quartic potential for each degree of freedom. In this way the double-well potential sufficient to entail the $\mathbb{Z}_2$-SBPT is generated. The occurrence of the $\mathbb{Z}_2$-SBPT dose not depend on the details of the constraint, which has to satisfy only the condition $V\rightarrow+\infty$ as $\phi_i\rightarrow\pm\infty$ for every $i=1,\cdots,N$. This characteristics have been pointed out in Refs. \cite{bc,b_3}. Unfortunately, the topological complication of the potential landscape of this model does not allow to highlight the connection with the revolution model with smooth potential introduced in this article. Next section will remedy this problem.

\subsection{A simplified version of the mean-field $\phi^4$ model}
\label{pphi4}

In Ref. \cite{b_5} a simplified version of the $\phi^4$ model has been introduced and studied in the mean-field version. The simplification is nothing more than the elimination of the quadratic term in the local potential. The new potential is therefore the following
\begin{equation}
V=\sum^{N}_{i=1}\frac{\phi_i^4}{4}-\frac{J}{2N}\left(\sum^{N}_{i=1}\phi_i\right)^2.
\label{Vpphi4}
\end{equation}
It has been shown that the quadratic term has no role in the generation of the $\mathbb{Z}_2$-SBPT which is identical to that of the traditional model apart quantitative differences. On the other hand, the quadratic term is a cause of great complication in the topological structure of the $\Sigma_{v,N}$'s which has been described in previous section. Thanks to this simplification, the potential landscape undergoes a topological trivialization, indeed, only three critical points survive against the immense multitude of the model with non-vanishing quadratic term. We speak of topological trivialization because to have a double-well potential at least two global minima with index $0$ and a central saddle point with index $1$ are needed.
\begin{figure}
	\begin{center}
		\includegraphics[width=0.235\textwidth]{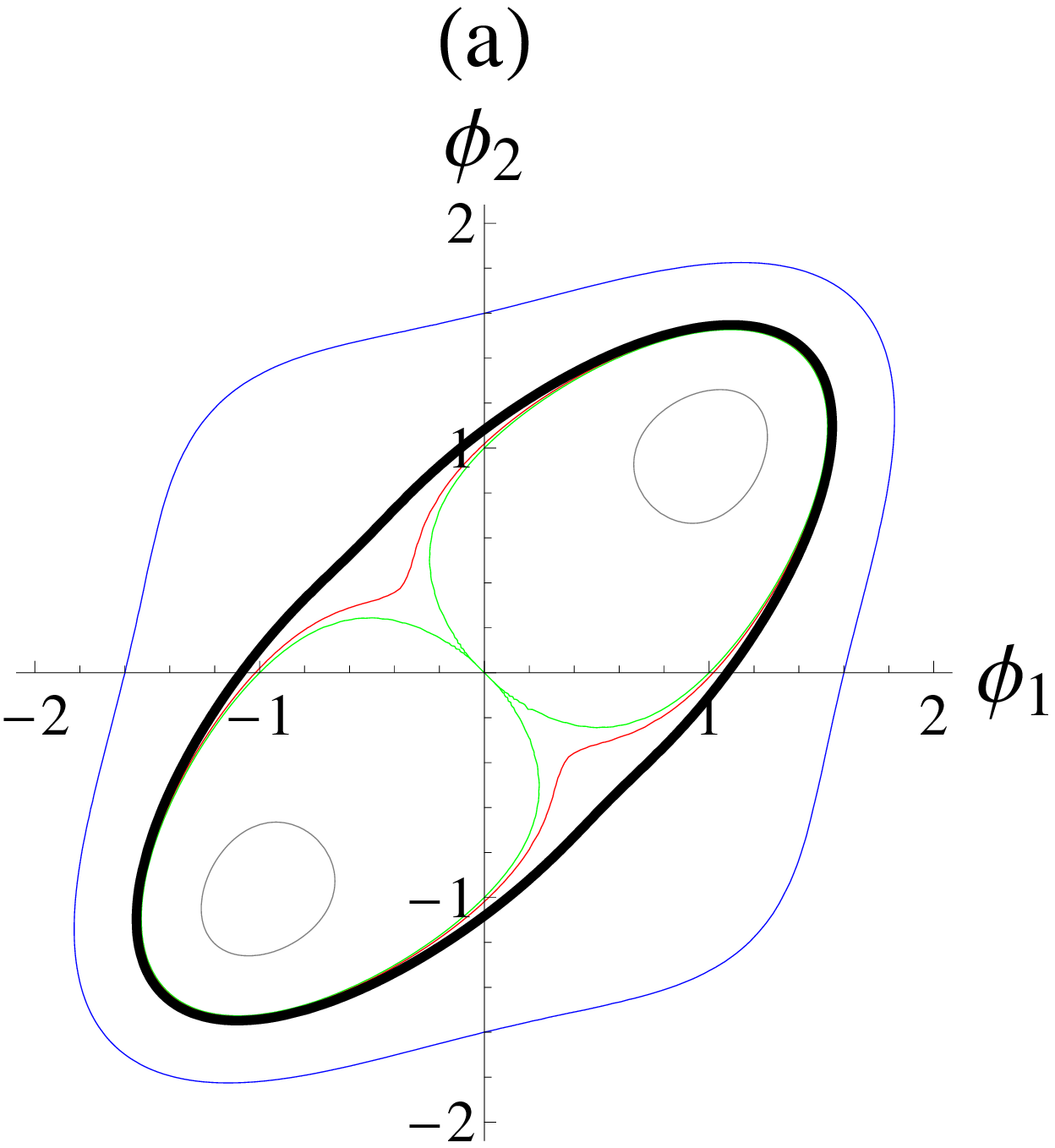}
		\includegraphics[width=0.235\textwidth]{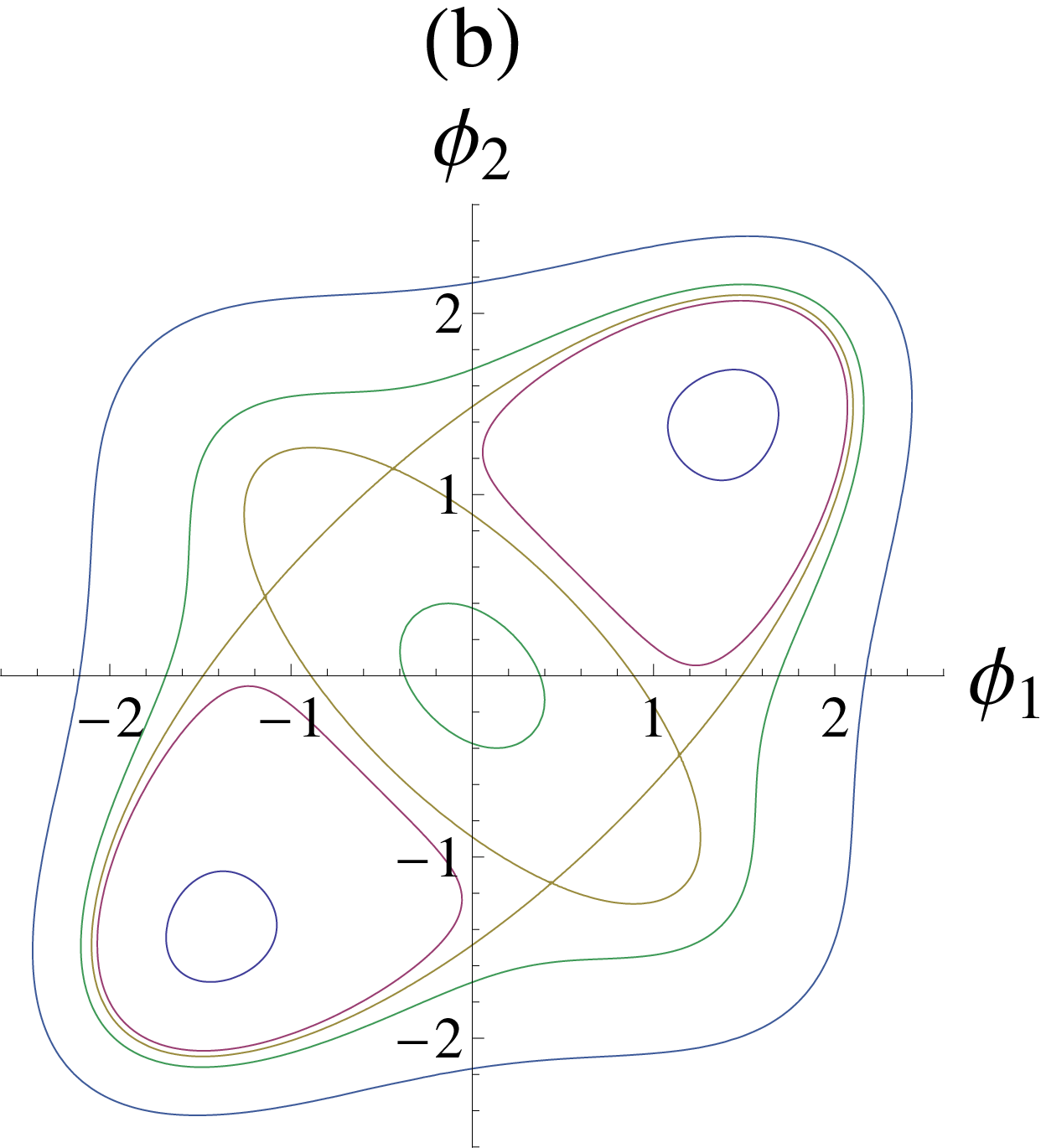}
		\caption{(a) Some $\Sigma_{v,N}$'s for $N=2$ of the mean-field  simplified $\phi^4$ model (\ref{Vpphi4}) for $J=1$ and $v=-0.4, ~0, ~0.01 ,~0.05,~ 1$. $\Sigma_{0.05,2}$ is marked. It is the boundary between the dumbbell-shaped $\Sigma_{v,N}$'s and those which are not. $0.05$ is only a numerical estimate because it is not possible to evaluate it analytically. (b) The same of the panel (a) for the mean-field $\phi^4$ model (\ref{Vphi4}) for  $J=1$ and $v=-1.8, ~-0.6, ~-0.4375 ,~-0.1,~ 2$.}
		\label{pphi4phi4}
	\end{center}
\end{figure}
In Fig. \ref{pphi4phi4} the proliferation of the critical points in the mean-field $\phi^4$ model due to the non-vanishing quadratic term of the local potential is already evident at $N=2$.

In the case of the model (\ref{Vpphi4}) the connection between the model with smooth potential proposed in this article and a physical model becomes truly evident. We want to emphasize that the potentials of the two models have the same topological and geometric structure, indeed they can be transformed one into the other by a mere variation of the shape. A comparison between the panel (a) of Fig. \ref{fig_revs_sigmav} and the panel (a) of Fig. \ref{pphi4phi4} clarifies this at $N=2$. The difference is that only in the case of the model (\ref{Vrevs}) the analytical calculation of the microcanonical entropy $s(v,m)$ is feasible. In the case of the mean-field simplified $\phi^4$ model the large deviation theory may be applied as it has been made in Ref. \cite{hk} for the traditional model.

\section{Conclusions}

In this paper we have introduced two Hamiltonian models with continuous $\mathbb{Z}_2$-SBPT entailed by some sufficiency conditions given on the potential energy landscape. These conditions are specified in Ref. \cite{bc,b_2}. The substantial feature is a double-well potential with gap proportional to the number of degree of freedom $N$. This implies that the factor $e^{-\beta V}$ competes with the density of states at constant magnetization entailing the critical temperature.

In Ref. \cite{b_3} it has been proven a straightforward theorem according to which dumbbell-shaped $\Sigma_{v,N}$'s are necessary and sufficiency condition to entail a $\mathbb{Z}_2$-SBPT. Roughly speaking, a $\Sigma_{v,N}$ is dumbbell-shaped if it is made up by two major components connected by a shrink neck. Generally, such a $\Sigma_{v,N}$ stems from a double-well potential, as it is the case of the models introduced here. In this framework the thermodynamic critical potential $\langle v\rangle_c$ results to be exactly in correspondence of a critical $\Sigma_{v_c,N}$ in the sense that it is the boundary between the dumbbell-shaped $\Sigma_{v_c,N}$'s for $v<v_c$ and those which are not for $v>v_c$. 

The model (\ref{Vrev}) here introduced is a toy model but it serves as a basis for introducing the model (\ref{Vrevs}) with smooth potential which instead has characteristics very similar to those of some physical models. Furthermore, the model (\ref{Vrevs}) can belongs to several universality classes besides the classical one by modulating its free parameters. As an example, the case of the short-range-2d Ising model has been studied. However, there is a difference that we are unable to account for: the specific heat has a logarithmic divergence in the case of the Ising model, while it has a jump in the model (\ref{Vrevs}).

At the end of the paper, the results for the mean-field $\phi^4$ model in Refs. \cite{aarz,b_1,bc,b_5} have been compared with those of the models here introduced in order to show the link with physical models. The main result lies in the fact that the potential landscape of a simplified version of the $\phi^4$ model introduced in Ref. \cite{b_5} is completely equivalent to that of the model (\ref{Vrevs}) with smooth potential from a geometric and topological point of view. This suggests that the framework proposed in Ref. \cite{b_3} to describe $\mathbb{Z}_2$-SBPTs by means of dumbbell-shaped $\Sigma_{v,N}$'s may be applicable for any physical model with continuous potential.


Finally, the model here introduced may be suitable also for didactic purposes and more analytical studies. In particular, we suggest the possibility of studying the curvature properties of the $\Sigma_{v,N}$'s of the model with smooth potential, e.g., by Gaussian curvature, to be related with the $\mathbb{Z}_2$-SBPT.

\begin{acknowledgments}
I would like to thank ................
\end{acknowledgments}

\end{document}